\documentclass[prd,aps,groupedaddress,showpacs,nofootinbib,amsmath]{revtex4}
\usepackage{epsf,epsfig,graphicx}

\newcounter{muni}

\begin{document}
\hbadness=10000 \pagenumbering{arabic}

\title{$\eta_c$ mixing effects on charmonium and $B$ meson decays}

\author{Yu-Dai Tsai$^{1,2}$}
\email{b94201020@ntu.edu.tw}
\author{Hsiang-nan Li$^{1,2,3}$}
\email{hnli@phys.sinica.edu.tw}
\author{Qiang Zhao$^{4,5}$}
\email{zhaoq@ihep.ac.cn }

\affiliation{$^{1}$Institute of Physics, Academia Sinica, Taipei,
Taiwan 115, Republic of China} \affiliation{$^{2}$Department of
Physics, Tsing-Hua University, Hsinchu, Taiwan 300, Republic of
China}
\affiliation{$^{3}$Department of Physics, National Cheng-Kung University,\\
Tainan, Taiwan 701, Republic of China}

\affiliation{$^4$ Institute of High Energy Physics, Chinese Academy
of Sciences, Beijing 100049, People’s Republic of China}

\affiliation{$^5$ Theoretical Physics Center for Science Facilities,
Chinese Academy of Sciences, Beijing 100049, People’s Republic of China}

\begin{abstract}

We include the $\eta_c$ meson into the $\eta$-$\eta'$-$G$ mixing
formalism constructed in our previous work, where $G$ represents the
pseudoscalar gluball. The mixing angles in this tetramixing matrix
are constrained by theoretical and experimental implications from
relevant hadronic processes. Especially, the angle between $\eta_c$
and $G$ is found to be about $11^\circ$ from the measured decay
widths of the $\eta_c$ meson. The pseudoscalar glueball mass $m_G$,
the pseudoscalar densities $m_{qq,ss,cc}$ and the $U(1)$ anomaly
matrix elements associated with the mixed states are solved from the
anomalous Ward identities. The solution $m_G\approx 1.4$ GeV
obtained from the $\eta$-$\eta'$-$G$ mixing is confirmed, while
$m_{qq}$ grows to above the pion mass, and thus increases
perturbative QCD predictions for the branching ratios
$Br(B\to\eta'K)$. We then analyze the $\eta_c$-mixing effects on
charmonium magnetic dipole transitions, and on the
$B\to\eta^{(\prime)}K_S$ branching ratios and CP asymmetries, which
further improve the consistency between theoretical predictions and
data. A predominant observation is that the $\eta_c$ mixing enhances
the perturbative QCD predictions for $Br(B\to\eta'K)$ by 18\%, but does not
alter those for $Br(B\to\eta K)$. The puzzle due to the large
$Br(B\to\eta'K)$ data is then resolved.

\bigskip

\end{abstract}


\maketitle

\section{INTRODUCTION}

It has been known that the gluonic and charm contents of the light
pseudoscalar mesons $\eta$ and $\eta'$ may have a crucial impact on
studies of many hadronic processes, such as $\eta^{(\prime)}$
electromagnetic (EM) transition form factors, $\eta^{(\prime)}$
radiative decays, and charmonium and heavy-flavor decays into
$\eta^{(\prime)}$. For a recent review on the gluonic effects in EM
transitions and in weak decays of charm and beauty hadrons, see
\cite{DiDonato:2011kr}. It remains a puzzle that theoretical
predictions for the $B\to\eta' K$ branching ratios are usually lower
than data, even after taking into account the $\eta$-$\eta'$ mixing
\cite{Kou:2001pm,Pham:2007nt}. Hence, it has been conjectured that
the gluonic content of the $\eta'$ meson plays a role in
accommodating the large branching ratios
\cite{Atwood:1997bn,Ahmady:1997fa,Eeg:2003pk,Du:1997hs,Du:1999us}. A
gluonic content of the $\eta^{(\prime)}$ meson has been inferred
from data of the radiative decays $P\to \gamma V$ and $V\to \gamma
P$ \cite{Ambrosino:2006gk,Escribano:2007cd} and from the charmonium
decays $J/\psi\to VP$ \cite{Li:2007ky,Escribano:2008rq}. However,
the gluonic contribution to the $B\to\eta^{(\prime)}$ transition
form factors was parametrized and tuned to fit data in the QCD
factorization approach \cite{Beneke:2002jn} and in the
soft-collinear effective theory \cite{Williamson:2006hb}, so no
conclusion on its importance could be drawn. This contribution was
calculated explicitly in the perturbative QCD (PQCD) approach
\cite{Hsu:2007qc} using the gluonic distribution amplitudes of the
$\eta^{(\prime)}$ meson from \cite{Ali:2003kg} and in QCD sum rules
\cite{Ball:2007hb}, and it was found to be small.

The charm content of the $\eta^{(\prime)}$ meson has been introduced
through the $\eta$-$\eta'$-$\eta_c$ mixing
\cite{Feldmann:1998vh,Feldmann:1998sh}. This formalism was extended
to the tetramixing among the $\pi$, $\eta$, $\eta'$, and $\eta_c$
mesons recently \cite{Peng:2011ue}, whose parameter set, including
the mixing angles and the hadronic parameters in the light-front
constituent quark model, was determined by a fit to data of relevant
meson transition form factors. An intermediate question is whether
this charm content affects the $B\to\eta^{(\prime)}K$ branching
ratios \cite{Halperin:1997as} and their CP asymmetries
\cite{Petrov:1997yf}, whose measurement might reveal new physics
signals. A potential deviation has been detected between the
mixing-induced CP asymmetries in the tree-dominated decays $B\to
J/\psi K_S$ and in the penguin-dominated decays $B\to\eta'K_S$.
Whether this deviation can be interpreted as a signal of new physics
depends on how large the tree pollution in the latter is. Though the
$\eta^{(\prime)}$-$\eta_c$ mixing is small
\cite{Ali:1997ex,Ahmady:1999tz}, there is lack of quantitative
estimate of its effect. It is then worthwhile to examine whether the
large $B\to\eta_c K$ amplitudes are able to compensate for the tiny
mixing, and to give a sizable impact on the $B\to\eta^{(\prime)}K$
decays.

The above subjects demand complete and precise understanding of the
gluonic and charm contents in the $\eta^{(\prime)}$ meson. We have
set up the $\eta$-$\eta'$-$G$ mixing in our previous work
\cite{Cheng:2008ss}, where $G$ denotes the physical pseudoscalar
glueball. This mixing was implemented into the equations of motion
from the anomalous Ward identities, that connect the vacuum to
$\eta, \eta'$, and $G$ transition matrix elements of the divergence
of axial-vector currents to the pseudoscalar densities and the
$U(1)$ anomaly. Solving these equations, the pseudoscalar glueball
mass $m_G$ was expressed in terms of phenomenological quantities
such as the $\eta, \eta'$ masses, the decay constants, and the
mixing angles. With the mixing angles measured from the $\phi \to
\gamma\eta, \gamma\eta'$ decays by KLOE \cite{Ambrosino:2006gk},
$m_G \approx 1.4$ GeV has been deduced, suggesting that the
$\eta(1405)$ meson is an interesting pseudoscalar glueball candidate
\cite{Cheng:2008ss,He:2009sb,Li:2009rk}. However, the solution for
the pseudoscalar density $m_{qq}$ associated with the $\eta_q$
component of the $\eta^{(\prime)}$ meson is still as low as its
conventional value $m_{qq}\approx m_\pi= 0.14$ GeV. It has been
postulated that a larger $m_{qq}\approx 0.2$ GeV can enhance the
$B\to\eta'$ transition form factors, and thus the $B\to\eta' K$
branching ratios significantly \cite{Akeroyd:2007fy}. Following this
vein, it was pointed out that the introduction of decay constants,
suppressed by the Okubo-Zweig-Iizuka (OZI) rule
\cite{OZI1,OZI2,OZI3}, into the equations from the anomalous Ward
identities can increase $m_{qq}$ \cite{Hsu:2007qc}.

We are motivated to formulate the tetramixing among $\eta$, $\eta'$,
$G$, and $\eta_c$, and to investigate its impact on charmonium and
$B$ meson decays in this paper. The mixing with the pion is not
considered here under the isospin symmetry. As a consequence of the
mixing, the $\eta_c$ meson contains a gluonic content, that modifies
the QCD calculation of its decay width. The fit to the observed
$\eta_c$ decay width determines the additional mixing angle
$\phi_Q\sim 11^\circ$ between $G$ and $\eta_c$. We shall explain
that the gluonic content of the $\eta_c$ meson further improves the
calculations for the decay widths of the charmonium magnetic dipole
transitions $J/\psi,\psi'\to\gamma\eta_c$ in association with the
unquenched mechanism proposed in \cite{Li:2007xr,Li:2011ss}, and
renders theoretical predictions in better agreement with data.
Together with the other mixing angles fixed in \cite{Cheng:2008ss},
we construct the tetramixing matrix, which implies the charm content
of the $\eta'$ meson consistent with that in \cite{Peng:2011ue}.
Implementing the tetramixing into the equations from the anomalous
Ward identities, we solve for the pseudoscalar glueball mass, the
pseudoscalar densities, and the $U(1)$ anomalies. It is found that
the inclusion of the $\eta_c$ mixing does not alter our prediction
for the pseudoscalar glueball mass, but increases the pseudoscalar
density $m_{qq}$ to above the pion mass, even in the absence of the
OZI-suppressed decay constants.

Moreover, the charm content of the $\eta^{(\prime)}$ meson allows
the $B\to\eta^{(\prime)}K$ decays via the $B\to\eta_cK$ channel.
Simply adopting the amplitudes evaluated in the PQCD approach at the
next-to-leading-order (NLO) accuracy \cite{Xiao:2008sw,Liu:2009kz},
we estimate the $\eta_c$ mixing effects on the
$B\to\eta^{(\prime)}K$ decays. It will be demonstrated that the
additional tree contribution from $B\to\eta_cK$ increases the
$B\to\eta'K$ branching ratios by about 18\%, but does not change the
$B\to\eta K$ branching ratios. Combining the mechanisms from the
larger $m_{qq}$ and the charm content, the puzzle due to the large
$B\to\eta' K$ branching ratios is resolved. On the other hand, the
charm content of the $\eta^{(\prime)}$ meson has minor influences on
the direct and mixing-induced CP asymmetries in the
$B\to\eta^{(\prime)}K_S$ decays. Nevertheless, we do see the
modification toward accommodating the deviation between the measured
mixing-induced CP asymmetries in the $B\to J/\psi K_S$ and
$B\to\eta'K_S$ decays. The $\eta_c$ mixing effects on the CP
asymmetries in the $B\to\eta K_S$ decays are also negligible.

We set up the $\eta$-$\eta'$-$G$-$\eta_c$ tetramixing formalism in
Sec.~II, which contains one more mixing angle $\phi_Q$ between $G$
and $\eta_c$ compared to the $\eta$-$\eta'$-$G$ mixing. The angle
$\phi_Q$ is then determined, and the resultant gluonic and charm
contents of the $\eta^{(\prime)}$ meson are compared with those
obtained in the literature. In Sec.~III we solve for the
pseudoscalar glueball mass, the pseudoscalar densities, and the
$U(1)$ anomalies appearing in the mixing formalism, and discuss
their phenomenological implications. The $\eta_c$ mixing effects on
charmonium magnetic dipole transitions and on the
$B\to\eta^{(\prime)}K$ decays are also investigated. Section IV
contains the summary and comments on other works, that present
observations different from ours.

\section{$\eta$-$\eta'$-$G$-$\eta_c$ MIXING}  \label{mixing}

In this section we formulate the $\eta$-$\eta'$-$G$-$\eta_c$
tetramixing, determine the involved mixing angles, and implement the
mixing into the equations of motion from the anomalous Ward
identities.

\subsection{Mixing matrix}

We combine the Feldmann-Kroll-Stech (FKS) formalism for the
$\eta$-$\eta'$-$\eta_c$ mixing
\cite{Feldmann:1998vh,Feldmann:1998sh} and for the
$\eta$-$\eta'$-$G$ mixing \cite{Cheng:2008ss}, in which the
conventional singlet-octet basis and the quark-flavor basis $q\bar
q\equiv (u\bar u+d\bar d)/\sqrt{2}$ and $s\bar s$
\cite{Schechter:1992iz}, labeled by $\eta_q$ and $\eta_s$,
respectively, were adopted. We further introduce the unmixed
glueball state $g$ and the unmixed heavy-quark state $\eta_Q$. Let
the matrix $U_{34}$ ($U_{14}$, $U_{12}$) represent a rotation with
the 3-4 (1-4, 1-2) plane fixed. It is natural to first mix those
singlet states $\eta_1$, $g$, and $\eta_Q$, and then mix $\eta_1$
and $\eta_8$ to form the physical states $\eta$, $\eta'$:
\begin{equation}\label{qs}
   \left( \begin{array}{c}
    |\eta\rangle \\ |\eta'\rangle\\|G\rangle\\|\eta_c\rangle
   \end{array} \right)
   = U_{34}(\theta)U_{14}(\phi_G)U_{12}(\phi_Q)
   \left( \begin{array}{c}
    |\eta_8\rangle \\ |\eta_1\rangle\\|g\rangle\\|\eta_Q\rangle
   \end{array} \right),
\end{equation}
with the rotational matrices
\begin{eqnarray}
& &U_{34}(\theta)=\left( \begin{array}{cccc}
    \cos\theta & -\sin\theta & 0 &0\\
    \sin\theta & \cos\theta &0 &0\\
    0 &0 &1 &0\\
    0 &0 &0 &1
   \end{array} \right),\;\;\;
U_{14}(\phi_G)=\left( \begin{array}{cccc}
    1 &0 &0 &0\\
    0 &\cos\phi_G & \sin\phi_G &0 \\
    0 &-\sin\phi_G & \cos\phi_G &0\\
    0 &0 &0 &1
   \end{array} \right),\nonumber\\
& &U_{12}(\phi_Q)=\left( \begin{array}{cccc}
    1 &0 &0 &0\\
    0 &1 &0 &0\\
    0 &0 &\cos\phi_Q & \sin\phi_Q  \\
    0 &0 &-\sin\phi_Q & \cos\phi_Q
   \end{array} \right).
\end{eqnarray}
That is, we assume that the octet state $\eta_8$ does not mix with
the glueball, and that the heavy-flavor state mixes with the
pseudoscalar glueball more dominantly than with the $\eta_1$ state.

The octet and singlet states are related to the flavor states via
\begin{equation}
\left( \begin{array}{c}
    |\eta_8\rangle \\ |\eta_1\rangle\\|g\rangle\\|\eta_Q\rangle
   \end{array} \right)
   = U_{34}(\theta_i)
\left( \begin{array}{c}
    |\eta_q\rangle \\ |\eta_s\rangle\\|g\rangle\\|\eta_Q\rangle
   \end{array} \right),
\end{equation}
where $\theta_i$ is the ideal mixing angle with
$\cos\theta_i=\sqrt{1/3}$ and $\sin\theta_i=\sqrt{2/3}$, i.e.,
$\theta_i=54.7^\circ$. The flavor states are then transformed into
the physical states through the mixing matrix
\begin{eqnarray}
U(\theta,\phi_G,\phi_Q)&=&U_{34}(\theta)U_{14}(\phi_G)U_{12}(\phi_Q)U_{34}(\theta_i),\nonumber\\
&=&\left(
\begin{array}{cccc}
c\theta c\theta_i-s\theta c\phi_G s\theta_i & -c\theta s\theta_i-s\theta c\phi_G c\theta_i
& -s\theta s\phi_G c\phi_Q& -s\theta s\phi_G s\phi_Q\\
s\theta c\theta_i+c\theta c\phi_G s\theta_i & -s\theta
s\theta_i+c\theta c\phi_G c\theta_i & c\theta s\phi_G c\phi_Q& c\theta s\phi_G s\phi_Q\\
-s\phi_G s\theta_i &-s\phi_G c\theta_i & c\phi_G
c\phi_Q & c\phi_G s\phi_Q \\
0 &0 & -s\phi_Q & c\phi_Q
   \end{array} \right),\label{mut}
\end{eqnarray}
with the notations $c\theta\equiv\cos\theta$ and
$s\theta\equiv\sin\theta$. Equation~(\ref{mut}) approaches the
mixing matrix in \cite{Cheng:2008ss} in the $\phi_Q\to 0$ limit, and
the $\eta$-$\eta'$ mixing matrix
\cite{Feldmann:1998vh,Feldmann:1998sh} in the $\phi_Q, \phi_G\to 0$
limit.

As stated in the Introduction, we have assumed isospin symmetry,
i.e. no mixing with $\pi^0$, and ignored other possible admixtures
from radial excitations. The decay constants $f_q$, $f_s$ and $f_c$
are defined via the matrix elements
\begin{eqnarray}
   \langle 0|\bar q\gamma^\mu\gamma_5 q|\eta_q(P)\rangle
   &=& -\frac{i}{\sqrt2}\,f_q\,P^\mu,\nonumber \\
   \langle 0|\bar s\gamma^\mu\gamma_5 s|\eta_s(P)\rangle
   &=& -i f_s\,P^\mu,\nonumber\\
   \langle 0|\bar c\gamma^\mu\gamma_5 c|\eta_Q(P)\rangle
   &=& -i f_c\,P^\mu,\label{deffq}
\end{eqnarray}
for the light quark $q=u$ or $d$. The other decay constants of the
$\eta_q$, $\eta_s$, and $\eta_Q$ mesons and of the unmixed glueball,
which are suppressed by the OZI rule, can be introduced in a similar
way \cite{Hsu:2007qc}:
\begin{eqnarray}
   & &\langle 0|\bar q\gamma^\mu\gamma_5 q|\eta_s(P),g(P),\eta_Q(P)\rangle
   = -\frac{i}{\sqrt2}\,f_{s,g,c}^q\,P^\mu , \nonumber\\
   & &\langle 0|\bar s\gamma^\mu\gamma_5 s|\eta_q(P),g(P),\eta_Q(P)\rangle
   = -i f_{q,g,c}^s\,P^\mu ,\nonumber\\
   & &\langle 0|\bar c\gamma^\mu\gamma_5 c|\eta_q(P),\eta_s(P),g(P)\rangle
   = -i f_{q,s,g}^c\,P^\mu .
\label{offdiagonal}
\end{eqnarray}
The decay constants associated with the $\eta$, $\eta'$, $G$, and
$\eta_c$ physical states in the following matrix elements
\begin{eqnarray}\label{deffh}
   & &\langle 0|\bar q\gamma^\mu\gamma_5 q|\eta(P),\eta^{\prime}(P),G(P),\eta_c(P)\rangle
   = -\frac{i}{\sqrt2}\,f_{\eta,\eta^{\prime},G,\eta_c}^q\,P^\mu
   ,\nonumber\\
   & &\langle 0|\bar s\gamma^\mu\gamma_5 s|\eta(P),\eta^{\prime}(P),G(P),\eta_c(P)\rangle
   = -i f_{\eta,\eta^{\prime},G,\eta_c}^s\,P^\mu ,\nonumber\\
   & &\langle 0|\bar c\gamma^\mu\gamma_5 c|\eta(P),\eta^{\prime}(P),G(P),\eta_c(P)\rangle
   = -i f_{\eta,\eta^{\prime},G,\eta_c}^c\,P^\mu ,
\end{eqnarray}
are related to those associated with the $\eta_q$, $\eta_s$, $g$,
$\eta_Q$ states through
\begin{eqnarray}
\left(
\begin{array}{ccc}
f_\eta^q & f_\eta^s & f_\eta^c \\
f_{\eta'}^q & f_{\eta'}^s & f_{\eta'}^c \\
f_G^q &f_G^s &f_G^c \\
f_{\eta_c}^q & f_{\eta_c}^s & f_{\eta_c}^c
\end{array} \right)=
U(\theta,\phi_G,\phi_Q) \left(
\begin{array}{ccc}
f_q & f_q^s & f_q^c\\
f_s^q & f_s & f_s^c \\
f_g^q & f_g^s & f_g^c \\
f_c^q & f_c^s & f_c
\end{array} \right).\label{fpi}
\end{eqnarray}

We sandwich the equations of motion for the anomalous Ward
identities
\begin{eqnarray}
   \partial_\mu(\bar q\gamma^\mu\gamma_5 q) &=& 2im_q\,\bar q\gamma_5 q
   +\frac{\alpha_s}{4\pi}\,G_{\mu\nu}\,\widetilde{G}^{\mu\nu},\nonumber\\
\partial_\mu(\bar s\gamma^\mu\gamma_5 s) &=& 2im_s\,\bar s\gamma_5s
   +\frac{\alpha_s}{4\pi}\,G_{\mu\nu}\,\widetilde{G}^{\mu\nu},\nonumber\\
\partial_\mu(\bar c\gamma^\mu\gamma_5 c) &=& 2im_c\,\bar c\gamma_5c
   +\frac{\alpha_s}{4\pi}\,G_{\mu\nu}\,\widetilde{G}^{\mu\nu},
   \label{eom}
\end{eqnarray}
between vacuum and $|\eta\rangle$, $|\eta'\rangle$, $|G\rangle$, and
$|\eta_c\rangle$, where $m_{q,s,c}$ are the quark masses,
$G_{\mu\nu}$ is the field-strength tensor and
$\widetilde{G}^{\mu\nu}$ is the dual field-strength tensor.
Following the procedure in \cite{Hsu:2007qc}, we derive
\begin{eqnarray}
M_{qsgc}=U^\dagger(\theta,\phi_G,\phi_Q) M^2
U(\theta,\phi_G,\phi_Q)\tilde J,\label{matrix}
\end{eqnarray}
in which the matrices are written as
\begin{eqnarray}
M_{qsgc}&=&\left(\begin{array}{ccc}
m_{qq}^2+\sqrt{2}G_q/f_q & m_{sq}^2+G_q/f_s & m_{cq}^2+G_q/f_c\\
m_{qs}^2+\sqrt{2}G_s/f_q & m_{ss}^2+G_s/f_s & m_{cs}^2+G_s/f_c\\
m_{qg}^2+\sqrt{2}G_g/f_q & m_{sg}^2+G_g/f_s & m_{cg}^2+G_g/f_c\\
m_{qc}^2+\sqrt{2}G_c/f_q & m_{sc}^2+G_c/f_s & m_{cc}^2+G_c/f_c\\
\end{array}\right),\label{maun}\\
M^2&=&\left(\begin{array}{cccc}
  m_{\eta}^2 & 0 &0 &0\\
  0 & m_{\eta'}^2&0 &0\\
  0 &0 & m_G^2 &0\\
  0 &0 &0 & m_{\eta_c}^2
\end{array} \right),\;\;\;\;
\tilde J=\left(\begin{array}{ccc}
   1 & f_q^s/f_s &f_q^c/f_c\\
  f_s^q/f_q & 1 & f_s^c/f_c \\
  f_g^q/f_q &f_g^s/f_s & f_g^c/f_c\\
  f_c^q/f_q &f_c^s/f_s & 1
\end{array} \right),\label{I}
\end{eqnarray}
with the $\eta$, $\eta'$, $G$, $\eta_c$ meson masses
$m_{\eta,\eta',G,\eta_c}$, and the abbreviations for the
pseudoscalar densities and the $U(1)$ anomaly matrix elements
\begin{eqnarray}
m_{qq,qs,qg,qc}^2&\equiv&\frac{\sqrt{2}}{f_q}\langle 0|m_u\bar u
i\gamma_5 u+m_d\bar d
i\gamma_5 d|\eta_q,\eta_s,g,\eta_Q\rangle\;,\nonumber\\
m_{sq,ss,sg,sc}^2&\equiv&\frac{2}{f_s}\langle 0|m_s\bar s i\gamma_5
s|\eta_q,\eta_s,g,\eta_Q\rangle, \nonumber\\
m_{cq,cs,cg,cc}^2&\equiv&\frac{2}{f_c}\langle 0|m_c\bar c i\gamma_5
c|\eta_q,\eta_s,g,\eta_Q\rangle, \nonumber\\
G_{q,s,g,c}&\equiv &\langle 0|\alpha_sG{\tilde
G}/(4\pi)|\eta_q,\eta_s,g,\eta_Q\rangle. \label{mqq}
\end{eqnarray}

In the following analysis we neglect all the OZI-suppressed decay
constants defined in Eq.~(\ref{offdiagonal}) according to the FKS
scheme. Equation~(\ref{I}) then leads to the following equations
explicitly
\begin{eqnarray}
M_{qsgc}^{11}&=&m_\eta^2(c\theta c\theta_i-s\theta c\phi_G s\theta_i)^2
+m_{\eta'}^2(s\theta c\theta_i+c\theta c\phi_G s\theta_i)^2
+m_G^2(s\phi_G s\theta_i)^2,\label{11}\\
M_{qsgc}^{12}&=&-m_\eta^2(c\theta c\theta_i-s\theta c\phi_G s\theta_i)
(c\theta s\theta_i+s\theta c\phi_G c\theta_i)\nonumber\\
& &+m_{\eta'}^2(s\theta c\theta_i+c\theta c\phi_G s\theta_i)(-s\theta
s\theta_i+c\theta c\phi_G c\theta_i)
+m_G^2(s\phi_G)^2c\theta_i s\theta_i,\label{12}\\
M_{qsgc}^{13}&=&-m_\eta^2(c\theta c\theta_i-s\theta c\phi_G s\theta_i)
s\theta s\phi_G s\phi_Q\nonumber\\
& &+m_{\eta'}^2(s\theta c\theta_i+c\theta c\phi_G s\theta_i)
c\theta s\phi_G s\phi_Q
-m_G^2 c\phi_Gs\phi_G s\theta_i s\phi_Q,\\
M_{qsgc}^{21}&=&M_{qsgc}^{12},\\
M_{qsgc}^{22}&=&m_\eta^2(c\theta s\theta_i+s\theta c\phi_G c\theta_i)^2+m_{\eta'}^2(-s\theta
s\theta_i+c\theta c\phi_G c\theta_i)^2
+m_G^2(s\phi_Gc\theta_i)^2,\\
M_{qsgc}^{23}&=&m_\eta^2(c\theta s\theta_i+s\theta c\phi_G c\theta_i)
s\theta s\phi_G s\phi_Q\nonumber\\
& &+m_{\eta'}^2(-s\theta s\theta_i+c\theta c\phi_G c\theta_i)
c\theta s\phi_G s\phi_Q
-m_G^2 c\phi_Gs\phi_G c\theta_i s\phi_Q,\\
M_{qsgc}^{31}&=&-m_\eta^2(c\theta c\theta_i-s\theta c\phi_G s\theta_i)
s\theta s\phi_G c\phi_Q\nonumber\\
& &+m_{\eta'}^2(s\theta c\theta_i+c\theta c\phi_G s\theta_i)
c\theta s\phi_G c\phi_Q
-m_G^2 c\phi_Gs\phi_G s\theta_i c\phi_Q,\label{m31}\\
M_{qsgc}^{32}&=&m_\eta^2(c\theta s\theta_i+s\theta c\phi_G c\theta_i)
s\theta s\phi_G c\phi_Q\nonumber\\
& &+m_{\eta'}^2(-s\theta s\theta_i+c\theta c\phi_G c\theta_i)
c\theta s\phi_G c\phi_Q
-m_G^2 c\phi_Gs\phi_G c\theta_i c\phi_Q,\label{m32}\\
M_{qsgc}^{33}&=&m_\eta^2(s\theta s\phi_G)^2 c\phi_Q s\phi_Q
+m_{\eta'}^2(c\theta s\phi_G)^2 c\phi_Q s\phi_Q\nonumber\\
& &+m_G^2 (c\phi_G)^2 c\phi_Q s\phi_Q-m_{\eta_c}^2c\phi_Q s\phi_Q,\\
M_{qsgc}^{41}&=&-m_\eta^2(c\theta c\theta_i-s\theta c\phi_G s\theta_i)
s\theta s\phi_G s\phi_Q\nonumber\\
& &+m_{\eta'}^2(s\theta c\theta_i+c\theta c\phi_G s\theta_i)
c\theta s\phi_G s\phi_Q
-m_G^2 c\phi_Gs\phi_G s\theta_i s\phi_Q,\label{m41}\\
M_{qsgc}^{42}&=&m_\eta^2(c\theta s\theta_i+s\theta c\phi_G
c\theta_i)s\theta s\phi_G s\phi_Q\nonumber\\
& &+m_{\eta'}^2(-s\theta s\theta_i+c\theta c\phi_G c\theta_i)
c\theta s\phi_G s\phi_Q
-m_G^2 c\phi_Gs\phi_G c\theta_i s\phi_Q,\label{m42}\\
M_{qsgc}^{43}&=&m_\eta^2(s\theta s\phi_G s\phi_Q)^2 +m_{\eta'}^2(
c\theta s\phi_G s\phi_Q)^2+m_G^2 (c\phi_G s\phi_Q)^2+m_{\eta_c}^2(c\phi_Q)^2.
\end{eqnarray}

\subsection{Mixing angles}

There is already extensive discussion on the determination of the
mixing angle $\theta$ in the literature, whose value still varies
in a finite range. For example, $-17^\circ < \theta < -11^\circ$ has
been extracted in \cite{Kou:1999tt}, assuming the presence of the
gluonic content in the $\eta'$ meson.
We choose $\theta=-11^\circ$, which corresponds to a sizable gluonic
content \cite{Kou:1999tt}, and is appropriate for our choice of
$\phi_G$ below. The value of $\phi_G$ varies in a
wide range, depending on the parametrization of the mixing matrix,
experimental inputs, and fitting procedures \cite{Cheng:2008ss}. Even
its central value can take a number between $10^\circ\lesssim \phi_G\lesssim
30^\circ$, such as $\phi_G=(12\pm
13)^\circ$ in \cite{Escribano:2007cd} and $\phi_G=(33\pm
13)^\circ$ in \cite{Escribano:2008rq}.
We take the value $\phi_G=12^\circ$, which was also considered in
\cite{Cheng:2008ss}.

The last angle $\phi_Q$ can be determined by the $\eta_c$ total
width and the $\eta_c\to\gamma\gamma$ decay width. As a $c\bar{c}(^1
S_0)$ state, the value of $\Gamma_{\rm tot}$ is quite large among
the charmonia below the $D\bar{D}$ threshold. During the past few
years, the experimental results for the $\eta_c$ total width and the
two-photon decay branching ratios vary drastically. The Particle
Data Group (PDG) 2008 \cite{Amsler:2008zb} listed $\Gamma_{\rm
tot}=26.7\pm 3.0$ MeV and $Br(\eta_c\to
\gamma\gamma)=(2.4^{+1.1}_{-0.9})\times 10^{-4} $, respectively. In
contrast, the PDG2010 \cite{Nakamura:2010zzi} presents $\Gamma_{\rm
tot}=28.6\pm 2.2$ MeV and $Br(\eta_c\to \gamma\gamma)=(6.3\pm
2.9)\times 10^{-5}$. The BESIII Collaboration measured the $\eta_c$
total width recently in $\psi'\to\gamma\eta_c$ and found
$\Gamma_{\rm tot}=32.0\pm 1.2\pm 1.0$ MeV
\cite{Collaboration:2011ii}. The interesting tendency is that the
total width of $\eta_c$ becomes broader than the previous
measurements. Such a change favors the scenario that the $\eta_c$
has a glueball content in association with the $c\bar{c}(^1 S_0)$
component as analyzed below. One also notices the much smaller
$Br(\eta_c\to\gamma\gamma)$ listed in the PDG2010 than in PDG2008,
which may be caused by the old total width data employed in the
extraction of the two-photon branching ratio. Since the averaged
partial decay widths for $\eta_c\to\gamma\gamma$ are unchanged in
PDG2008 and PDG2010, i.e.
$\Gamma(\eta_c\to\gamma\gamma)=6.7^{+0.9}_{-0.8}$ keV, it might be
more appropriate to adopt the experimental data for the two-photon
partial width instead of for the branching ratio in the following
analysis.


The strong decay of $c\bar{c}(^1 S_0)$ annihilation via two-gluon
radiation can be related to its two-photon decay. To lowest order,
one has
\begin{equation}
\frac{\Gamma(^1 S_0\to gg)}{\Gamma(^1 S_0\to\gamma\gamma)}\simeq
\frac{2\alpha_s^2}{9e_c^4\alpha_e^2}
=\frac{9}{8}\left(\frac{\alpha_s}{\alpha_e}\right)^2,
\end{equation}
where $e_c=2/3$ is the charge of the $c$ quark, and $\alpha_s$ and
$\alpha_e$ are the strong and EM coupling constants, respectively.
With  $\Gamma(^1 S_0\to gg)\simeq \Gamma_{\rm tot}=28.6\pm 2.2$ MeV
from PDG2010 \cite{Nakamura:2010zzi} and $\Gamma(^1
S_0\to\gamma\gamma)=6.7^{+0.9}_{-0.8}$ keV, we extract
$\alpha_s\simeq 0.41\sim 0.49$ ($\alpha_s\simeq 0.72\sim 1.2$ if one
adopts the branching ratio $Br(\eta_c\to \gamma\gamma)=(6.3\pm
2.9)\times 10^{-5}$ from PDG2010 \cite{Nakamura:2010zzi}),
which is much larger than the running coupling $\alpha_s(m_c)\simeq
0.24\sim 0.26$ \cite{Nakamura:2010zzi}. Even with the first-order
QCD correction \cite{Kwong:1987ak} taken into account:
\begin{equation}
\frac{\Gamma(^1 S_0\to gg)}{\Gamma(^1 S_0\to\gamma\gamma)}\simeq
\frac{9}{8}\left(\frac{\alpha_s}{\alpha_e}\right)^2
\frac{(1+4.8\frac{\alpha_s}{\pi})}{(1-3.4\frac{\alpha_s}{\pi})},
\end{equation}
we still have larger values $\alpha_s\simeq 0.28\sim 0.33$.

As defined in Eq.~(\ref{mut}), the $\eta_c$ meson has the wave
function
\begin{equation}\label{etac-wave}
|\eta_c\rangle = -s\phi_Q |g\rangle +c\phi_Q |\eta_Q\rangle .
\end{equation}
The strong decay amplitude can be parametrized as
\begin{eqnarray}
\langle gg|\hat{V}|\eta_c\rangle &= &  -s\phi_Q \langle
gg|\hat{V}| g\rangle +c\phi_Q \langle gg|\hat{V}|\eta_Q\rangle
\nonumber\\
&= & \left(-\frac{\pi s\phi_Q}{\alpha_s}+c\phi_Q\right) \langle
gg|\hat{V}|\eta_Q\rangle ,
\end{eqnarray}
where $\hat{V}$ denotes the potential for annihilating the
$c\bar{c}$ states and the glueball into two gluons. Since the
glueball does not pay a price for coupling to the $|gg\rangle$
state, we have the ratio $\langle gg|\hat{V}| g\rangle/ \langle
gg|\hat{V}|\eta_Q\rangle =\pi/\alpha_s$ given by the empirical gluon
power counting \cite{Close:2005vf}. Therefore, the mixing provides a
correction to the strong decays different from the correction to the
EM coupling $\alpha_e^2 \to (c\phi_Q)^2\alpha_e^2$. Consequently,
one has, to the leading QCD correction,
\begin{equation}
\frac{\Gamma(\eta_c\to gg)}{\Gamma(\eta_c\to\gamma\gamma)}
=\frac{9}{8}\frac{(-\pi
t\phi_Q+\alpha_s)^2}{\alpha_e^2}\frac{(1+4.8\frac{\alpha_s}{\pi})}
{(1-3.4\frac{\alpha_s}{\pi})},
\end{equation}
with $t\phi_Q\equiv \tan \phi_Q$. Requiring
$\alpha_s=\alpha_s(m_c)\simeq 0.25$, we deduce $\phi_Q\simeq
-1.4^\circ$ or $10.4^\circ$ ($\phi_Q\simeq -6.9^\circ$ or
$15.6^\circ$, if $Br(\eta_c\to\gamma\gamma)= (6.3\pm 2.9)\times
10^{-5}$ is adopted). If inserting the recent BESIII value
$\Gamma_{\rm tot}=32.0\pm 1.2\pm 1.0$ MeV
\cite{Collaboration:2011ii}, we extract $\phi_Q\simeq -1.7^\circ$ or
$10.8^\circ$. It is noted in advance that the data of the radiative
decays $J/\psi, \psi^\prime\to \gamma\eta_c$ favor the positive
solution as illustrated in Sec.~III C. Based on the above
determinations of $\phi_Q$, we adopt $\phi_Q=11^\circ$ in this work
for the purpose of estimation.

We should mention a recent calculation of the $^1S_0$ decays into
light hadrons and two photons in the framework of nonrelativistic
QCD (NRQCD) \cite{Guo:2011tz}, in which the $O(\alpha_s v^2)$
corrections were found to be crucial for accommodating the observed
$\eta_c\to gg$ and $\gamma\gamma$ widths. The involved long-distance
matrix elements for the $\eta_c$ meson still suffer large
uncertainties, and their values were obtained by data fitting in
\cite{Guo:2011tz}. Hence, it is a fair comment that the present data
cannot distinguish the NRQCD results with the $O(\alpha_s v^2)$
corrections from the mixing scenario with a small glueball component
in the $\eta_c$ meson. Meanwhile, it should be realized that the
presence of the glueball component would affect the total and EM
decay widths of the $\eta_c$ meson differently, and thus make an
impact on their ratio at leading order.

With the angles $\theta=-11^\circ$, $\phi_G=12^\circ$ and
$\phi_Q=11^\circ$, the mixing matrix in Eq.~(\ref{mut}) is
explicitly written as
\begin{eqnarray}
U &=&\left(
\begin{array}{cccc}
0.720  & -0.693& 0.039  & 0.008\\
0.673  & 0.711  & 0.200     &0.039\\
-0.170 &-0.120  & 0.960     & 0.186 \\
0         &0         & -0.191   & 0.982
   \end{array} \right).\label{mutv}
\end{eqnarray}
Compared to the parametrization for the $\eta$-$\eta'$-$\eta_c$
mixing \cite{Feldmann:1998vh,Feldmann:1998sh}, the role of the small
angle $\theta_c=-1^\circ\pm 0.1^\circ$ has been played by our
$\phi_G\phi_Q$. The matrix for the $\eta$-$\eta'$-$\eta_c$ mixing is
given by \cite{Feldmann:1998vh}
\begin{eqnarray}
|\eta\rangle&=&0.77|\eta_q\rangle-0.63|\eta_s\rangle-0.006|\eta_Q\rangle,\nonumber\\
|\eta'\rangle&=&0.63|\eta_q\rangle+0.77|\eta_s\rangle-0.016|\eta_Q\rangle,\nonumber\\
|\eta_c\rangle&=&0.015|\eta_q\rangle+0.008|\eta_s\rangle+|\eta_Q\rangle,\label{mutf}
\end{eqnarray}
in which the charm content in the $\eta^{(\prime)}$ meson has a sign
opposite to that in Eq.~(\ref{mutv}). The fit to the data of
$\Gamma(J/\psi\to\gamma\eta')/\Gamma(J/\psi\to\gamma\eta_c)$
\cite{Feldmann:1998vh} actually cannot discriminate the sign of these
coefficients. Another difference appears in the
$\eta_{q,s}$ components of the $\eta_c$ meson. Since these are small
components, there is no inconsistency between the results in
Eqs.~(\ref{mutv}) and (\ref{mutf}).

The charm contents of the $\eta$ and $\eta'$ mesons
have the same sign in the FKS scheme.
This feature differs from the parametrization for the
$\pi$-$\eta$-$\eta'$-$\eta_c$ tetramixing based on the group
decomposition $SO(4)=SO(3)\otimes SO(3)$ \cite{Peng:2011ue},
\begin{eqnarray}
U_{\pi\eta\eta'\eta_c} &=&\left(
\begin{array}{cccc}
0.9895 & 0.0552 & -0.1119& 0.0342\\
-0.1082 & 0.8175 & -0.5614& -0.0259\\
0.0590 &0.5696 & 0.8160 & 0.0452 \\
-0.0395 &-0.0065 & -0.0478 & 0.9960
   \end{array} \right),\label{mutt}
\end{eqnarray}
where the charm contents of the $\eta$ and $\eta'$ mesons are
opposite in sign.  A careful look reveals that these matrix elements
are small due to the destruction of large numbers, so they are
sensitive to experimental inputs. Varying the inputs slightly, one
could get the matrix elements of the same sign in
\cite{Peng:2011ue}. The $\eta$ meson has a pion component $-0.11$,
while the $\eta'$ meson has a smaller pion component 0.06 as
indicated in Eq.~(\ref{mutt}). Therefore, it is appropriate to
compare the charm contents of the $\eta'$ meson in Eqs.~(\ref{mutv})
and (\ref{mutt}), both of which are about 0.04.

\section{$\eta_c$-MIXING EFFECTS}

In this section we solve Eq.~(\ref{matrix}) from the anomalous Ward
identities, and then investigate the phenomenological impacts from
the $\eta_c$ mixing on charmonium magnetic dipole transitions and on
the $B\to\eta^{(\prime)}K$ decays.

\subsection{Decay constants}

To obtain the decay constants defined in Eq.~(\ref{fpi}), we need
the inputs of $f_q$, $f_s$ and $f_c$ for the flavor eigenstates. The
value of $f_q$ is close to the pion decay constant $f_\pi$, such as
$f_q=(1.07\pm 0.02)f_\pi$ extracted in \cite{Feldmann:1998vh}. Our
analysis indicates that the variation of $f_q$ has almost no
influence, so we simply set it to $f_q=f_\pi=131$ MeV. The value of
$f_s$ is more uncertain, for which the result $f_s/f_q\approx
1.2\sim 1.3$ also from \cite{Feldmann:1998vh} is employed. Since the
ratio $f_s/f_q$ gives a more significant effect, we shall examine
how our outcomes depend on its variation. For $f_c$, we adopt the
value $f_c=487.4$ MeV \cite{Peng:2011ue}. Ignoring the terms
suppressed by the OZI rule, Eq.~(\ref{fpi}) leads to
\begin{eqnarray}
\left(
\begin{array}{ccc}
f_\eta^q & f_\eta^s & f_\eta^c \\
f_{\eta'}^q & f_{\eta'}^s & f_{\eta'}^c \\
f_G^q &f_G^s &f_G^c \\
f_{\eta_c}^q & f_{\eta_c}^s & f_{\eta_c}^c
\end{array} \right)=
U\left(
\begin{array}{ccc}
f_q & 0 & 0\\
0 & f_s & 0 \\
0 & 0 & 0 \\
0 & 0 & f_c
\end{array} \right)=\left(
\begin{array}{ccc}
113     & -90.8     & 3.69\\
106     & 93.1  & 19.0 \\
-26.7   & -15.7     & 91.0\\
0       & 0     & 478
\end{array} \right)
.\label{fpi1}
\end{eqnarray}

The larger decay constants $f_{\eta^{(\prime)}}^{q,s}$ are close to
those appearing in the literature \cite{Feldmann:1998vh}, and in the
perturbative calculations of the $B\to\eta^{(\prime)}K$ decays
\cite{Charng:2006zj,Xiao:2008sw,Beneke:2002jn}. It is natural that
$f_{\eta_c}^c$ is almost the same as $f_c$. Those decay constants
$f_G^{q,s,c}$ may have phenomenological applications, after the
pseudoscalar glueball is identified. Because of $\theta= -11^\circ$,
we have the ratio $f_\eta^c/f_{\eta'}^c=-\tan\theta=0.24$, which is
similar to $-0.006/(-0.016)=0.38$ in
\cite{Feldmann:1998vh,Feldmann:1998sh}, but has an opposite sign to
$-0.0259/0.0452=-0.57$ in \cite{Peng:2011ue}. Note that the sign of
$f_{\eta'}^c$ is still not certain, which was found to be positive
in \cite{Halperin:1997as,Yuan:1997ts}, but negative in
\cite{Feldmann:1998vh,Feldmann:1998sh}. According to
\cite{Wu:2011gf}, a positive $f_{\eta^{(\prime)}}^c$ would enhance
the $\eta^{(\prime)}$ transition form factor in the light-cone PQCD
approach. Nevertheless, the enhancement will be compensated by
varying other parameters, such as the constituent quark masses, so
that the corresponding data can still be accommodated in their
theoretical framework \cite{Wu:2011gf}. Our result
$f_{\eta'}^c=19.0$ MeV, arising from the phenomenological
determination, is approximately equal to 21.9 MeV in
\cite{Peng:2011ue}, and lower than 50-180 MeV computed from QCD low
energy theorem \cite{Halperin:1997as}. The similarity to the value
in \cite{Peng:2011ue} is nontrivial, since we have constructed the
mixing matrix via the angles $\theta$, $\phi_G$ and $\phi_Q$, which
were determined in a different way. It has been conjectured
\cite{Franz:2000ee} that the value of $f_{\eta'}^c$ was
overestimated in \cite{Halperin:1997as}. Hence, we disagree with
their speculation that the charm content of the $\eta'$ meson alone
can exhaust the large $B\to\eta' K$ branching ratios.

Our result $f_{\eta'}^c=19.0$ MeV is larger than $-(6.3\pm 0.6)$ MeV
in magnitude in \cite{Feldmann:1998vh}, and about 8 times larger
than 2.4 MeV extracted from the data of
$Br(J/\psi\to\gamma\eta')/Br(J/\psi\to\gamma\eta_c)$
\cite{Ahmady:1999tz}.  As stated above, the data of
$Br(J/\psi\to\gamma\eta')/Br(J/\psi\to\gamma\eta_c)$ cannot fix the
sign of $f_{\eta'}^c$ in \cite{Feldmann:1998vh}. Besides, the
smaller value in \cite{Feldmann:1998vh} is attributed to the tiny
$\eta_c$ mixing, which is a consequence of the anomalous Ward
identities in Eq.~(\ref{eom}) without the OZI-suppressed terms.
Including the OZI-suppressed terms, a larger range for $f_{\eta'}^c$
is expected. The analysis in \cite{Feldmann:1998vh,Ahmady:1999tz},
together with that in \cite{Ali:1997nh,Feldmann:2000hs} which also
concluded a small $\eta'$-$\eta_c$ mixing, were performed in the
$\eta$-$\eta'$-$\eta_c$ mixing formalism. That is, only a single
channel $J/\psi \to \gamma\eta_Q \to \gamma\eta'$ contributes to
$Br(J/\psi\to\gamma\eta')$, which is perhaps too strong of an
assumption. With the more general $\eta$-$\eta'$-$G$-$\eta_c$
tetramixing, an additional channel $J/\psi \to \gamma g \to
\gamma\eta'$ exists. For the purpose of illustration, we regard
$\eta(1405)$ as the pseudoscalar glueball \cite{Cheng:2008ss}, and
assign a finite fraction of the $J/\psi \to \gamma \eta(1405/1475)$
amplitude to $J/\psi \to \gamma\eta(1405)$. Employing the data in
PDG2010 \cite{Nakamura:2010zzi}, it is easy to find that the above
two channels, multiplied by the mixing matrix elements 0.039 and 0.2
in Eq.~(\ref{mutv}), respectively, are comparable to each other. If
a destructive interference occurs between the two channels, a larger
$\eta'$-$\eta_c$ mixing will be allowed. On the other hand, some
theoretical estimates on $f_{\eta'}^c$ also implied a small
$\eta'$-$\eta_c$ mixing
\cite{Beneke:2002jn,Franz:2000ee,Ali:1997ex}, in which the
annihilation of the $c\bar c$ pair into two gluons, followed by
their combination into the $\eta'$ meson, was considered. Strictly
speaking, what they investigated is the extrinsic charm content of
the $\eta'$ meson, while the $\eta'$-$\eta_c$ mixing results in the
intrinsic charm content \cite{Peng:2011ue}. Therefore, their
observation does not contradict to ours, and supports that the
intrinsic charm content analyzed in this work might be more
relevant.

\subsection{Glueball mass, Pseudoscalar densities, and $U(1)$ anomalies}

Given the decay constants $f_q$, $f_s$ and $f_c$, we solve
Eq.~(\ref{matrix}) in order to get an idea of the magnitude of the
pseudoscalar gluball mass, the pseudoscalar densities, and the
$U(1)$ anomaly matrix elements. There are 12 equations from the
$4\times 3$ matrix equation in Eq.~(\ref{matrix}). Even having
dropped all the OZI-suppressed decay constants, there are still too
many unknowns listed above. Hence, we intend to drop the
OZI-suppressed pseudoscalar densities, except $m_{cq}^2$, $m_{cs}^2$
and $m_{cg}^2$, whose values can serve as a check of their scaling
behavior in the large $N_c$ limit \cite{Cheng:2008ss}, $N_c$ being
the number of colors. Then one of Eqs.~(\ref{m31}), (\ref{m32}),
(\ref{m41}), and (\ref{m42}) becomes redundant, since both the ratio
of Eq.~(\ref{m31}) over Eq.~(\ref{m32}), and the ratio of
Eq.~(\ref{m41}) over Eq.~(\ref{m42}) give
\begin{eqnarray}
\frac{-m_\eta^2(c\theta c\theta_i-s\theta c\phi_G s\theta_i) s\theta
+m_{\eta'}^2(s\theta c\theta_i+c\theta c\phi_G s\theta_i) c\theta
-m_G^2 c\phi_G s\theta_i}{m_\eta^2(c\theta s\theta_i+s\theta c\phi_G
c\theta_i) s\theta +m_{\eta'}^2(-s\theta s\theta_i+c\theta c\phi_G
c\theta_i) c\theta-m_G^2 c\phi_G
c\theta_i}=\frac{\sqrt{2}f_s}{f_q},\label{rat}
\end{eqnarray}
which is identical to the formula derived in \cite{Cheng:2008ss}.
Eventually, we solve for 11 unknowns, which include the pseudoscalar
glueball mass $m_G$, 6 pseudoscalar densities $m_{qq}^2$,
$m_{ss}^2$, $m_{cq}^2$, $m_{cs}^2$, $m_{cg}^2$, and $m_{cc}^2$, and
4 $U(1)$ anomaly matrix elements $G_q$, $G_s$, $G_g$, and $G_c$.

According to Eq.~(\ref{11}), $m_{qq}^2$ is expressed as the
difference between the right-hand side of Eq.~(\ref{11}) and the
anomaly matrix element $G_q/f_q$. Equation~(\ref{12}) implies that
$G_q$ is proportional to $f_s$, so the above difference strongly
depends on the ratio $f_s/f_q$ \cite{Hsu:2007qc}.
Equation~(\ref{rat}) shows that the inclusion of the $\eta_c$ mixing
does not affect much the solution $m_G\approx 1.4$ GeV
\cite{Cheng:2008ss}. The value of $m_G$ is insensitive to the
uncertain mixing angle $\phi_G$ also, because of $c\phi_G\approx
1$ for a small $\phi_G$. Therefore, $m_G$ is only sensitive to the
ratio $f_s/f_q$, and we shall consider the variation of this ratio
below. The solutions corresponding to $f_s/f_q=1.2$ and 1.3 are
collected in Table~\ref{tab:1}. It is seen that only $m_G$ and
$m_{qq}^2$ depend on the ratio $f_s/f_q$, and other quantities are
relatively stable. The results of $m_G$, $m_{ss}^2$, $G_q$, $G_s$,
and $G_g$ are similar to those from the $\eta$-$\eta'$-$G$ mixing
formalism \cite{Cheng:2008ss}. Namely, $m_{ss}^2$ in
Table~\ref{tab:1} respects the leading $N_c$ relation
$m_{ss}^2=2m_K^2-m_\pi^2$, and we do not observe the enhancement
claimed in \cite{Gerard:2006ch}. Note that a larger pseudoscalar
glueball mass ($>2$ GeV) has been postulated in a dynamical analysis
of the mixing in the pseudoscalar channel \cite{Mathieu:2009sg}.
However, if their assumption on the meson couplings is relaxed, a
lower mass can be attained. Our solutions respect the hierarchy
$|m_{cc}^2|\gg |m_{cg}^2|\gg |m_{cq,cs}^2|$ in the large $N_c$
limit. The values of $m_{cc}^2$ are consistent with the relation
$m_{cc}^2\approx m_{\eta_c}^2$ \cite{Feldmann:1998vh}, but the
magnitude of $G_c$ is a bit smaller than that in
\cite{Feldmann:1998vh}.

It has been shown that the $B\to\eta'$ transition form factors and
the $B\to\eta' K$ branching ratios are sensitive to the pseudoscalar
density $m_{qq}$ \cite{Akeroyd:2007fy}, which defines the
normalization of the two-parton twist-3 distribution amplitudes for
the $\eta_q$ state. Its value is usually assumed to be the pion
mass, $m_{qq}\approx m_\pi$, with which the branching ratios
$Br(B^\pm\to\eta' K^\pm)\approx 51\times 10^{-6}$ and
$Br(B^0\to\eta' K^0)\approx 50\times 10^{-6}$ have been obtained in
NLO PQCD \cite{Xiao:2008sw}. These results are still lower than the
data $Br(B^\pm\to\eta' K^\pm)\approx (71.1\pm 2.6)\times 10^{-6}$
and $Br(B^0\to\eta' K^0)\approx (66.1\pm 3.1)\times 10^{-6}$
\cite{HFAG}. It was then demonstrated that $m_{qq}$ can be increased
up to 0.2 GeV by introducing the OZI-suppressed decay constants
$f_q^s$ and $f_s^q$ into the equation for the $\eta$-$\eta'$ mixing
\cite{Hsu:2007qc}. With the enhanced $m_{qq}$, it is likely to
explain the large $Br(B\to\eta' K)$ data with larger $B\to\eta'$
transition form factors. Since $f_q^s$ and $f_s^q$ are free
parameters, whether $m_{qq}$ can reach 0.2 GeV is not conclusive.
Table~\ref{tab:1} indicates that the $\eta_c$ mixing, which receives
a phenomenological support as elucidated in Sec.~II B, can enlarge
$m_{qq}^2$ for $f_s/f_q=1.3$ by a factor 2, from $m_{qq}^2\approx
0.012$ GeV$^2$ in Eq.~(29) of \cite{Cheng:2008ss} to 0.023 GeV$^2$.
Note that the values of the mixing angle $\theta$ for Eq.~(29) in
\cite{Cheng:2008ss} and for Table~\ref{tab:1} are different.
However, $m_{qq}^2$ does not much depend on the variation of
$\theta$.

\begin{table}[t]
\caption{Solutions corresponding to $f_s/f_q=1.2$ and $f_s/f_q=1.3$.}
\begin{tabular}{cc|c|cccccc|c c c c} \toprule
$f_q$ &$f_s/f_q$ & $m_G$ (GeV) & $m_{qq}^2$ & $m_{ss}^2$
& $m_{cq}^2$ & $m_{cs}^2$ & $m_{cg}^2$ & $m_{cc}^2$ (GeV$^2$)
& $G_q$ & $G_s$ & $G_g$ & $G_c$ (GeV$^3$) \\
\colrule
$f_\pi$ &           $1.2$ &       $1.519$ &
$0.067$ &       $0.443$&    $-0.156$ &
$-0.092$ &      $-1.197$ &  $8.648 $ &        $0.053$ &
$0.031$  &          $-0.023$ &  $-0.004$    \\
\colrule
$f_\pi$     & $1.3$     & $1.376$ &
$0.023$  & $0.457$ & $-0.149$ &
$-0.081$ & $-1.283$ & $8.631$ & $0.056$ &
$0.030$  & $-0.016$ & $-0.003$    \\
\botrule
\end{tabular} \label{tab:1}
\end{table}

\subsection{Charmonium M1 transitions}

There has been a long-standing puzzle from the magnetic dipole (M1)
transition $J/\psi\to \gamma\eta_c$. In contrast to the success of
the nonrelativistic potential models in offering an overall good
description of the charmonium spectrum, the predicted M1 transitions
between the vector and pseudoscalar charmonia appear to have
significant discrepancies. Namely, the predicted partial decay width
$\Gamma^{NR}(J/\psi\to\gamma\eta_c)\simeq 2.4\sim 2.9$ keV
\cite{Godfrey:1985xj,Barnes:2005pb} is obviously larger than the
experimental data in PDG2010~\cite{Nakamura:2010zzi},
$Br(J/\psi\to\gamma\eta_c)=(1.7\pm 0.4)\%$, i.e.,
$\Gamma(J/\psi\to\gamma\eta_c)=(1.58\pm 0.37)$ keV. The PDG2010
value is mainly weighted by the CLEO data $Br(J/\psi\to
\gamma\eta_c)=(1.98\pm 0.09\pm 0.30)\%$ \cite{:2008fb}. Although the
CLEO data bring the experimental and theoretical values closer, the
discrepancy remains nontrivial after taking into account another
channel $\psi^\prime\to\gamma\eta_c$: the potential models predicted
$\Gamma^{NR}(\psi^\prime\to\gamma\eta_c)\simeq 9.6\sim 9.7$ keV,
i.e., $Br(\psi^\prime\to\gamma\eta_c)\simeq 3.2\%$, while the CLEO
measurement gives $Br(\psi^\prime\to\gamma\eta_c)=(4.32\pm 0.16\pm
0.60)\times 10^{-3}$ \cite{:2008fb}.

Theoretical efforts of studying the charmonium EM M1 transitions in
the framework of nonrelativistic multipole expansions can be found
in \cite{Eichten:1974af,Eichten:1975ag,Eichten:1978tg,
Eichten:1979ms,Ebert:2002pp,Eichten:2007qx,Brambilla:2005zw,Brambilla:2004wf}.
Recently, a nonrelativistic effective field theory was applied to
$J/\psi\to \gamma\eta_c$ \cite{Brambilla:2005zw}, in which the
radiative decay width $\Gamma(J/\psi\to \gamma\eta_c)=(1.5\pm 1.0)$
keV up to correction of $O(v_c^2/m_c^2)$ was obtained with a rather
large uncertainty. This approach becomes unreliable in $\psi^\prime
\to \gamma\eta_c$, because the $c\bar{c}$ pair cannot be treated as
a weakly bound system anymore. The lattice QCD calculations of these
processes were reported in \cite{Dudek:2006ej,Dudek:2009kk}. In the
quenched approximation the result for $J/\psi\to \gamma\eta_c$
turns out to be in agreement with the potential models, while that
for $\psi^\prime\to\gamma\eta_c$ is much smaller and compatible with
the data within uncertainties. That is, the quenched lattice QCD
does not resolve the puzzle completely.

A possible resolution arises from the accommodation of open
threshold effects as an unquenched mechanism in the charmonium M1
transitions \cite{Li:2007xr,Li:2011ss}. Because of the presence of the
open $D\bar{D}$ threshold, the transition
$\psi^\prime\to\gamma\eta_c$ would experience more influence from
the $D\bar{D}$ threshold, which naturally lowers the partial decay
width predicted by the potential models. In contrast, the open
threshold effects on $J/\psi\to\gamma\eta_c$ are relatively small,
since the mass of $J/\psi$ is located rather far away from the
$D\bar{D}$ threshold. However, the uncertainties from the
unquenched effects are significant as shown in
\cite{Li:2007xr,Li:2011ss}, so there is still room for the
glueball-$\eta_c$ mixing mechanism in the present experimental and
theoretical situations.

Starting with Eq.~(\ref{etac-wave}), we express the
quenched M1 transition amplitude as
\begin{eqnarray}
\langle\eta_c|H_{em}|J/\psi,\psi'\rangle
&=& -s\phi_Q\langle g|H_{em}|J/\psi,\psi'\rangle +
c\phi_Q\langle\eta_Q|H_{em}|J/\psi,\psi'\rangle \nonumber\\
&\simeq & (-s\phi_Q\frac{\alpha_s}{\pi}+c\phi_Q)
\langle\eta_Q|H_{em}|J/\psi,\psi'\rangle \ ,
\end{eqnarray}
where $\langle\eta_Q|H_{em}|J/\psi,\psi'\rangle$ is equivalent to
the potential-model M1 transition amplitude for the $J/\psi$,
$\psi'$ mesons, and the gluon counting rule has been implemented in
the second line. One immediately notices that in order to lower the
M1 transition partial width, a positive $\phi_Q$ is 
favored. Given $\phi_Q\simeq 11^\circ$, the quenched M1
transition partial widths are lowered by about $7 \%$. 
That is, the glueball-$\eta_c$ mixing does improve the overall
consistency between the potential-model predictions and the data for
the charmonium M1 transitions.

\subsection{$B^0 \rightarrow \eta^{(\prime)}K_S$ decays}

As stated in the Introduction, a potential deviation has been
detected between the mixing-induced CP asymmetries in the
tree-dominated decays $B\to J/\psi K_S$ and in the penguin-dominated
decays $B\to\eta'K_S$, which demands a deeper theoretical
understanding. Besides, the branching ratios of the $B\to\eta' K$
decays predicted in the PQCD approach up to NLO are still much lower
than the data. The value of $f_{\eta'}^c$ obtained in Sec.~III A
suggests a quantitative reexamination of the tree contribution from
$B\to\eta_c K$ to the direct and mixing-induced CP asymmetries and
the branching ratios of the $B^0\to\eta^{(\prime)}K_S$ decays. The
$\lambda_{CP}$ factor for this study is defined as
\begin{eqnarray}
\lambda_{CP}
=\eta_fe^{-2i\beta}\frac{\langle f\vert H_{eff}\vert \overline{B^0}\rangle}
{\langle f\vert H_{eff}\vert B^0\rangle}\;,
\end{eqnarray}
with the eigenvalue $\eta_f=-1$ and the weak phase $\beta$. The
direct and mixing-induced CP asymmetries are then derived from
\begin{eqnarray}
A_{CP}^{dir}
=\frac{|\lambda|^2-1}{1+|\lambda|^2},\;\;\;\;
A_{CP}^{mix}
=\frac{2Im{\lambda}}{1+|\lambda|^2}.
\end{eqnarray}

For the $B^0 \to \eta'K_S$ decays without the $\eta_c$ mixing, one has
\begin{eqnarray}
\lambda_{CP}
&=&
\eta_fe^{-2i\beta}\frac{V_{ub}V_{us}^*T_{\eta'K}-V_{tb}V_{ts}^*P_{\eta'K}}
{V_{ub}^*V_{us}T_{\eta'K}-V_{tb}^*V_{ts}P_{\eta'K}},\label{lam}
\end{eqnarray}
where $V$'s are the CKM matrix elements, and $T$ and $P$ represent
the tree and penguin amplitudes, respectively. These decay
amplitudes have been evaluated up to NLO in the PQCD approach
\cite{Xiao:2008sw}, which lead to $A_{CP}^{dir}=0.024$ and
$A_{CP}^{mix}=0.667$\footnote{The quoted value of $A_{CP}^{mix}$
differs from that presented in \cite{Xiao:2008sw}, since the input
of the weak phase $\beta$ has been corrected.}. The tree amplitude
$T_{\eta_cK}$ has been also calculated in NLO PQCD
\cite{Liu:2009kz}, which gives the branching ratio $Br(B^0 \to
\eta_c K^0)=5.5\times10^{-4}$. Because the $B$ meson decay constants
$f_B=0.21$ GeV and $f_B=0.19$ GeV were adopted in \cite{Xiao:2008sw}
and \cite{Liu:2009kz}, respectively, and the $\eta_c$ meson decay
constants $f_{\eta_c}=0.478$ GeV were derived in Eq.~(\ref{fpi1}), and
$f_{\eta_c}=0.42$ GeV was adopted in \cite{Liu:2009kz}, we multiply
$T_{\eta_cK}$ in \cite{Liu:2009kz} by a factor
$(0.21/0.19)(0.478/0.42)=1.26$ for consistency. The increased NLO
PQCD prediction $Br(B^0 \to \eta_c K^0)=8.7\times10^{-4}$ then
agrees well with the data $(8.7\pm 1.9)\times10^{-4}$
\cite{HFAG}. Note that the relative strong
phases among the above amplitudes $T_{\eta'K}$, $P_{\eta'K}$ and
$T_{\eta_cK}$ are known in the PQCD approach, so we do not encounter
the difficulty mentioned in \cite{Petrov:1997yf}, and can derive the
modified CP asymmetries without ambiguity. The inclusion of the
$B\to\eta_c K$ channel into Eq.~(\ref{lam}),
\begin{eqnarray}
\lambda_{CP}&=& \eta_fe^{-i2\beta} \frac{V_{ub}V_{us}^*T_{\eta'K}
-V_{tb}V_{ts}^*P_{\eta'K} +c\theta s\phi_G s\phi_Q
V_{cb}V_{cs}^*T_{\eta_cK}}{V_{ub}^*V_{us}T_{\eta'K}
-V_{tb}^*V_{ts}P_{\eta'K} +c\theta s\phi_G s\phi_Q
V_{cb}^*V_{cs}T_{\eta_cK}},
\end{eqnarray}
yields $A_{CP}^{dir}=0.023$ and $A_{CP}^{mix}=0.664$. It is seen
that the $\eta_c$ mixing causes a negligible effect on the CP
asymmetries with $A_{CP}^{mix}$ moving slightly toward the central
value of the data, $0.59\pm 0.07$ \cite{HFAG}. The result of
$A_{CP}^{dir}$ is consistent with the data $0.05\pm 0.05$
\cite{HFAG}. Nevertheless, the $\eta_c$ mixing brings the branching
ratio $Br(B^0 \to \eta^{\prime} K^0)=50\times10^{-6}$ in NLO PQCD
\cite{Liu:2009kz} to $59\times10^{-6}$, which becomes closer to the
data $(66.1\pm 3.1)\times 10^{-6}$ \cite{HFAG}. Note that the
enhancement of the above branching ratio due to the charm content of
the $\eta'$ meson is larger than few percents estimated in
\cite{Gerard:2004je}.

Similarly, we investigate the impact of the $\eta_c$ mixing on the
$B^0\to \eta K_S$ decays. Without the charm content of the $\eta$
meson, the NLO PQCD analysis gave $A_{CP}^{dir}=-0.128$ and
$A_{CP}^{mix}= 0.659$ \cite{Xiao:2008sw}. The $\eta_c$ mixing then
modifies the above values into $A_{CP}^{dir}=-0.123$ and
$A_{CP}^{mix}=0.644$, namely, with a negligible effect. The
branching ratio $Br(B^0\to \eta K^0)$, becoming $2.3\times 10^{-6}$
from $2.1\times 10^{-6}$, is almost not changed by the $\eta_c$
mixing. The result is a bit higher than the data $Br(B^0\to\eta
K^0)=(1.12^{+0.30}_{-0.28})\times 10^{-6}$ \cite{HFAG}. However, if
using $\theta=-11^\circ$ in the present work, the destructive
interference for the $B\to\eta K$ decays from the $\eta$-$\eta'$
mixing would be stronger, which will lower their branching ratios.

\section{SUMMARY}

In this paper we have extended the $\eta$-$\eta'$-$G$ mixing
formalism constructed in our previous work to accommodate the
$\eta_c$ meson in a tetramixing scheme. The additional mixing angle
between $G$ and $\eta_c$ was determined to be about $11^\circ$ from
the observed widths of the $\eta_c$ meson decays into light hadrons
and $\gamma\gamma$. This mixing would have a leading impact on the
$\eta_c\to gg$ width instead of on the $\eta_c\to\gamma\gamma$ one,
such that the $O(\alpha_s v^2)$ corrections in NRQCD
\cite{Guo:2011tz} can be parametrized out. More precise measurement
of the $\eta_c$ total decay width and its decays into $\gamma\gamma$
can provide better constraints on the mixing scheme. Our tetramixing
matrix was found to be consistent with that constructed from the
$SO(3)\otimes SO(3)$ parametrization with a fit to data of relevant
transition form factors. Contrary to general opinions in the
literature, the present work suggests an reexamination of effects
from the charm content of the $\eta^{(\prime)}$ meson on the
$B\to\eta^{(\prime)}K$ decays.

We have shown that such a tetramixing scheme does increase the
pseudoscalar density $m_{qq}$ to above the pion mass, which thus
enhances theoretical predictions for the $B\to\eta'$ transition form
factors and the $B\to\eta'K$ branching ratios \cite{Akeroyd:2007fy}.
It has been observed that the charm content of the $\eta'$ meson
provides 18\% enhancement of the $B\to\eta' K$ branching ratios. The
combined mechanisms can push the predicted values in NLO PQCD to the
data easily, so the puzzle due to the large $Br(B\to\eta' K)$ is
resolved. With this work, we postulate that the charm content of the
$\eta'$ meson plays a more important role than the gluonic content
in accommodating the large $Br(B\to\eta' K)$. Nevertheless, the
$\eta_c$ mixing has negligible effects on the direct and
mixing-induced CP asymmetries of the $B\to\eta^{(\prime)}K_S$
decays, though the mixing-induced CP asymmetries move slightly
toward the central value of the data.

We have also investigated the impact of the tetramixing on the
present theoretical and experimental observations of charmonium
magnetic dipole transitions, and similar improvement is also seen:
the gluonic content of the $\eta_c$ meson decreases the decay widths
$\Gamma(J/\psi,\psi'\to\gamma\eta_c)$ predicted by the
nonrelativistic potential models by 7\%. It has been confirmed that
the $\eta_c$ mixing does not modify the prediction in
\cite{Cheng:2008ss} for the pseudoscalar glueball mass in the
vicinity of 1.4-1.5 GeV. This result makes the $\eta(1405)$ meson an
interesting candidate for the pseudoscalar glueball. However, one
should be aware of the complexity of underlying dynamics, such as
the octet-glueball coupling \cite{Mathieu:2009sg} and intermediate
meson rescattering in this mass region \cite{Wu:2011yx}, which
certainly affect the determination of the pseudoscalar glueball
mass.


\acknowledgments{

We acknowledge X. Liu, Z.J. Xiao, and Z.Q. Zhang for providing the
$B$ meson decay amplitudes in NLO PQCD. HNL and YDT thank H.Y. Cheng
for useful discussions, and the hospitality of the Institute of
High Energy Physics at the Chinese Academy of Sciences, where part of
this work was done. This work was supported, in part, by National
Science Council of R.O.C. under Grant No. NSC 98-2112-M-001-015-MY3,
the National Natural Science Foundation of China (Grant No. 11035006),
the Chinese Academy of Sciences (KJCX2-EW-N01), and the Ministry of Science
and Technology of China (2009CB825200). }

\bibliography{Reference_Hadron}%

\begin{thebibliography}{67}
\expandafter\ifx\csname natexlab\endcsname\relax\def\natexlab#1{#1}\fi
\expandafter\ifx\csname bibnamefont\endcsname\relax
  \def\bibnamefont#1{#1}\fi
\expandafter\ifx\csname bibfnamefont\endcsname\relax
  \def\bibfnamefont#1{#1}\fi
\expandafter\ifx\csname citenamefont\endcsname\relax
  \def\citenamefont#1{#1}\fi
\expandafter\ifx\csname url\endcsname\relax
  \def\url#1{\texttt{#1}}\fi
\expandafter\ifx\csname urlprefix\endcsname\relax\def\urlprefix{URL }\fi
\providecommand{\bibinfo}[2]{#2}
\providecommand{\eprint}[2][]{\url{#2}}

\bibitem[{\citenamefont{Di~Donato et~al.}(2012)\citenamefont{Di~Donato,
  Ricciardi, and Bigi}}]{DiDonato:2011kr}
\bibinfo{author}{\bibfnamefont{C.}~\bibnamefont{Di~Donato}},
  \bibinfo{author}{\bibfnamefont{G.}~\bibnamefont{Ricciardi}},
  \bibnamefont{and} \bibinfo{author}{\bibfnamefont{I.}~\bibnamefont{Bigi}},
  \bibinfo{journal}{Phys.Rev.} \textbf{\bibinfo{volume}{D85}},
  \bibinfo{pages}{013016} (\bibinfo{year}{2012}), \bibinfo{note}{18 pages, 2
  figures}, \eprint{1105.3557}.

\bibitem[{\citenamefont{Kou and Sanda}(2002)}]{Kou:2001pm}
\bibinfo{author}{\bibfnamefont{E.}~\bibnamefont{Kou}} \bibnamefont{and}
  \bibinfo{author}{\bibfnamefont{A.~I.} \bibnamefont{Sanda}},
  \bibinfo{journal}{Phys. Lett.} \textbf{\bibinfo{volume}{B525}},
  \bibinfo{pages}{240} (\bibinfo{year}{2002}), \eprint{hep-ph/0106159}.

\bibitem[{\citenamefont{Pham}(2008)}]{Pham:2007nt}
\bibinfo{author}{\bibfnamefont{T.~N.} \bibnamefont{Pham}},
  \bibinfo{journal}{Phys. Rev.} \textbf{\bibinfo{volume}{D77}},
  \bibinfo{pages}{014024} (\bibinfo{year}{2008}),
  \bibinfo{note}{[Erratum-ibid.D77:019905,2008]}, \eprint{0710.2412}.

\bibitem[{\citenamefont{Atwood and Soni}(1997)}]{Atwood:1997bn}
\bibinfo{author}{\bibfnamefont{D.}~\bibnamefont{Atwood}} \bibnamefont{and}
  \bibinfo{author}{\bibfnamefont{A.}~\bibnamefont{Soni}},
  \bibinfo{journal}{Phys.Lett.} \textbf{\bibinfo{volume}{B405}},
  \bibinfo{pages}{150} (\bibinfo{year}{1997}), \eprint{hep-ph/9704357}.

\bibitem[{\citenamefont{Ahmady et~al.}(1998)\citenamefont{Ahmady, Kou, and
  Sugamoto}}]{Ahmady:1997fa}
\bibinfo{author}{\bibfnamefont{M.~R.} \bibnamefont{Ahmady}},
  \bibinfo{author}{\bibfnamefont{E.}~\bibnamefont{Kou}}, \bibnamefont{and}
  \bibinfo{author}{\bibfnamefont{A.}~\bibnamefont{Sugamoto}},
  \bibinfo{journal}{Phys.Rev.} \textbf{\bibinfo{volume}{D58}},
  \bibinfo{pages}{014015} (\bibinfo{year}{1998}), \eprint{hep-ph/9710509}.

\bibitem[{\citenamefont{Eeg et~al.}(2003)\citenamefont{Eeg, Kumericki, and
  Picek}}]{Eeg:2003pk}
\bibinfo{author}{\bibfnamefont{J.~O.} \bibnamefont{Eeg}},
  \bibinfo{author}{\bibfnamefont{K.}~\bibnamefont{Kumericki}},
  \bibnamefont{and} \bibinfo{author}{\bibfnamefont{I.}~\bibnamefont{Picek}},
  \bibinfo{journal}{Phys. Lett.} \textbf{\bibinfo{volume}{B563}},
  \bibinfo{pages}{87} (\bibinfo{year}{2003}), \eprint{hep-ph/0304274}.

\bibitem[{\citenamefont{Du et~al.}(1998)\citenamefont{Du, Kim, and
  Yang}}]{Du:1997hs}
\bibinfo{author}{\bibfnamefont{D.-s.} \bibnamefont{Du}},
  \bibinfo{author}{\bibfnamefont{C.~S.} \bibnamefont{Kim}}, \bibnamefont{and}
  \bibinfo{author}{\bibfnamefont{Y.-d.} \bibnamefont{Yang}},
  \bibinfo{journal}{Phys. Lett.} \textbf{\bibinfo{volume}{B426}},
  \bibinfo{pages}{133} (\bibinfo{year}{1998}), \eprint{hep-ph/9711428}.

\bibitem[{\citenamefont{Du et~al.}(2002)\citenamefont{Du, Yang, and
  Zhu}}]{Du:1999us}
\bibinfo{author}{\bibfnamefont{D.-s.} \bibnamefont{Du}},
  \bibinfo{author}{\bibfnamefont{D.-s.} \bibnamefont{Yang}}, \bibnamefont{and}
  \bibinfo{author}{\bibfnamefont{G.-h.} \bibnamefont{Zhu}},
  \bibinfo{journal}{High Energy Phys.Nucl.Phys.} \textbf{\bibinfo{volume}{26}},
  \bibinfo{pages}{1} (\bibinfo{year}{2002}), \eprint{hep-ph/9912201}.

\bibitem[{\citenamefont{Ambrosino et~al.}(2007)}]{Ambrosino:2006gk}
\bibinfo{author}{\bibfnamefont{F.}~\bibnamefont{Ambrosino}}
  \bibnamefont{et~al.} (\bibinfo{collaboration}{KLOE}), \bibinfo{journal}{Phys.
  Lett.} \textbf{\bibinfo{volume}{B648}}, \bibinfo{pages}{267}
  (\bibinfo{year}{2007}), \eprint{hep-ex/0612029}.

\bibitem[{\citenamefont{Escribano and Nadal}(2007)}]{Escribano:2007cd}
\bibinfo{author}{\bibfnamefont{R.}~\bibnamefont{Escribano}} \bibnamefont{and}
  \bibinfo{author}{\bibfnamefont{J.}~\bibnamefont{Nadal}},
  \bibinfo{journal}{JHEP} \textbf{\bibinfo{volume}{05}}, \bibinfo{pages}{006}
  (\bibinfo{year}{2007}), \eprint{hep-ph/0703187}.

\bibitem[{\citenamefont{Li et~al.}(2008)\citenamefont{Li, Zhao, and
  Chang}}]{Li:2007ky}
\bibinfo{author}{\bibfnamefont{G.}~\bibnamefont{Li}},
  \bibinfo{author}{\bibfnamefont{Q.}~\bibnamefont{Zhao}}, \bibnamefont{and}
  \bibinfo{author}{\bibfnamefont{C.-H.} \bibnamefont{Chang}},
  \bibinfo{journal}{J. Phys.} \textbf{\bibinfo{volume}{G35}},
  \bibinfo{pages}{055002} (\bibinfo{year}{2008}), \eprint{hep-ph/0701020}.

\bibitem[{\citenamefont{Escribano}(2010)}]{Escribano:2008rq}
\bibinfo{author}{\bibfnamefont{R.}~\bibnamefont{Escribano}},
  \bibinfo{journal}{Eur. Phys. J.} \textbf{\bibinfo{volume}{C65}},
  \bibinfo{pages}{467} (\bibinfo{year}{2010}), \eprint{0807.4201}.

\bibitem[{\citenamefont{Beneke and Neubert}(2003)}]{Beneke:2002jn}
\bibinfo{author}{\bibfnamefont{M.}~\bibnamefont{Beneke}} \bibnamefont{and}
  \bibinfo{author}{\bibfnamefont{M.}~\bibnamefont{Neubert}},
  \bibinfo{journal}{Nucl. Phys.} \textbf{\bibinfo{volume}{B651}},
  \bibinfo{pages}{225} (\bibinfo{year}{2003}), \eprint{hep-ph/0210085}.

\bibitem[{\citenamefont{Williamson and Zupan}(2006)}]{Williamson:2006hb}
\bibinfo{author}{\bibfnamefont{A.~R.} \bibnamefont{Williamson}}
  \bibnamefont{and} \bibinfo{author}{\bibfnamefont{J.}~\bibnamefont{Zupan}},
  \bibinfo{journal}{Phys. Rev.} \textbf{\bibinfo{volume}{D74}},
  \bibinfo{pages}{014003} (\bibinfo{year}{2006}),
  \bibinfo{note}{[Erratum-ibid.D74:03901,2006]}, \eprint{hep-ph/0601214}.

\bibitem[{\citenamefont{Hsu et~al.}(2008)\citenamefont{Hsu, Charng, and
  Li}}]{Hsu:2007qc}
\bibinfo{author}{\bibfnamefont{J.-F.} \bibnamefont{Hsu}},
  \bibinfo{author}{\bibfnamefont{Y.-Y.} \bibnamefont{Charng}},
  \bibnamefont{and} \bibinfo{author}{\bibfnamefont{H.-n.} \bibnamefont{Li}},
  \bibinfo{journal}{Phys.Rev.} \textbf{\bibinfo{volume}{D78}},
  \bibinfo{pages}{014020} (\bibinfo{year}{2008}), \eprint{0711.4987}.

\bibitem[{\citenamefont{Ali and Parkhomenko}(2003)}]{Ali:2003kg}
\bibinfo{author}{\bibfnamefont{A.}~\bibnamefont{Ali}} \bibnamefont{and}
  \bibinfo{author}{\bibfnamefont{A.~Y.} \bibnamefont{Parkhomenko}},
  \bibinfo{journal}{Eur. Phys. J.} \textbf{\bibinfo{volume}{C30}},
  \bibinfo{pages}{367} (\bibinfo{year}{2003}), \eprint{hep-ph/0307092}.

\bibitem[{\citenamefont{Ball and Jones}(2007)}]{Ball:2007hb}
\bibinfo{author}{\bibfnamefont{P.}~\bibnamefont{Ball}} \bibnamefont{and}
  \bibinfo{author}{\bibfnamefont{G.~W.} \bibnamefont{Jones}},
  \bibinfo{journal}{JHEP} \textbf{\bibinfo{volume}{08}}, \bibinfo{pages}{025}
  (\bibinfo{year}{2007}), \eprint{0706.3628}.

\bibitem[{\citenamefont{Feldmann et~al.}(1998)\citenamefont{Feldmann, Kroll,
  and Stech}}]{Feldmann:1998vh}
\bibinfo{author}{\bibfnamefont{T.}~\bibnamefont{Feldmann}},
  \bibinfo{author}{\bibfnamefont{P.}~\bibnamefont{Kroll}}, \bibnamefont{and}
  \bibinfo{author}{\bibfnamefont{B.}~\bibnamefont{Stech}},
  \bibinfo{journal}{Phys. Rev.} \textbf{\bibinfo{volume}{D58}},
  \bibinfo{pages}{114006} (\bibinfo{year}{1998}), \eprint{hep-ph/9802409}.

\bibitem[{\citenamefont{Feldmann et~al.}(1999)\citenamefont{Feldmann, Kroll,
  and Stech}}]{Feldmann:1998sh}
\bibinfo{author}{\bibfnamefont{T.}~\bibnamefont{Feldmann}},
  \bibinfo{author}{\bibfnamefont{P.}~\bibnamefont{Kroll}}, \bibnamefont{and}
  \bibinfo{author}{\bibfnamefont{B.}~\bibnamefont{Stech}},
  \bibinfo{journal}{Phys. Lett.} \textbf{\bibinfo{volume}{B449}},
  \bibinfo{pages}{339} (\bibinfo{year}{1999}), \eprint{hep-ph/9812269}.

\bibitem[{\citenamefont{Peng and Ma}(2011)}]{Peng:2011ue}
\bibinfo{author}{\bibfnamefont{T.}~\bibnamefont{Peng}} \bibnamefont{and}
  \bibinfo{author}{\bibfnamefont{B.-Q.} \bibnamefont{Ma}},
  \bibinfo{journal}{Phys. Rev.} \textbf{\bibinfo{volume}{D84}},
  \bibinfo{pages}{034003} (\bibinfo{year}{2011}), \eprint{1107.5088}.

\bibitem[{\citenamefont{Halperin and Zhitnitsky}(1997)}]{Halperin:1997as}
\bibinfo{author}{\bibfnamefont{I.~E.} \bibnamefont{Halperin}} \bibnamefont{and}
  \bibinfo{author}{\bibfnamefont{A.}~\bibnamefont{Zhitnitsky}},
  \bibinfo{journal}{Phys. Rev.} \textbf{\bibinfo{volume}{D56}},
  \bibinfo{pages}{7247} (\bibinfo{year}{1997}), \eprint{hep-ph/9704412}.

\bibitem[{\citenamefont{Petrov}(1998)}]{Petrov:1997yf}
\bibinfo{author}{\bibfnamefont{A.~A.} \bibnamefont{Petrov}},
  \bibinfo{journal}{Phys.Rev.} \textbf{\bibinfo{volume}{D58}},
  \bibinfo{pages}{054004} (\bibinfo{year}{1998}), \eprint{hep-ph/9712497}.

\bibitem[{\citenamefont{Ali et~al.}(1998)\citenamefont{Ali, Chay, Greub, and
  Ko}}]{Ali:1997ex}
\bibinfo{author}{\bibfnamefont{A.}~\bibnamefont{Ali}},
  \bibinfo{author}{\bibfnamefont{J.}~\bibnamefont{Chay}},
  \bibinfo{author}{\bibfnamefont{C.}~\bibnamefont{Greub}}, \bibnamefont{and}
  \bibinfo{author}{\bibfnamefont{P.}~\bibnamefont{Ko}}, \bibinfo{journal}{Phys.
  Lett.} \textbf{\bibinfo{volume}{B424}}, \bibinfo{pages}{161}
  (\bibinfo{year}{1998}), \eprint{hep-ph/9712372}.

\bibitem[{\citenamefont{Ahmady and Kou}(1999)}]{Ahmady:1999tz}
\bibinfo{author}{\bibfnamefont{M.~R.} \bibnamefont{Ahmady}} \bibnamefont{and}
  \bibinfo{author}{\bibfnamefont{E.}~\bibnamefont{Kou}} (\bibinfo{year}{1999}),
  \eprint{hep-ph/9903335}.

\bibitem[{\citenamefont{Cheng et~al.}(2009)\citenamefont{Cheng, Li, and
  Liu}}]{Cheng:2008ss}
\bibinfo{author}{\bibfnamefont{H.-Y.} \bibnamefont{Cheng}},
  \bibinfo{author}{\bibfnamefont{H.-n.} \bibnamefont{Li}}, \bibnamefont{and}
  \bibinfo{author}{\bibfnamefont{K.-F.} \bibnamefont{Liu}},
  \bibinfo{journal}{Phys. Rev.} \textbf{\bibinfo{volume}{D79}},
  \bibinfo{pages}{014024} (\bibinfo{year}{2009}), \eprint{0811.2577}.

\bibitem[{\citenamefont{He et~al.}(2010)\citenamefont{He, Huang, and
  Yan}}]{He:2009sb}
\bibinfo{author}{\bibfnamefont{S.}~\bibnamefont{He}},
  \bibinfo{author}{\bibfnamefont{M.}~\bibnamefont{Huang}}, \bibnamefont{and}
  \bibinfo{author}{\bibfnamefont{Q.-S.} \bibnamefont{Yan}},
  \bibinfo{journal}{Phys.Rev.} \textbf{\bibinfo{volume}{D81}},
  \bibinfo{pages}{014003} (\bibinfo{year}{2010}), \eprint{0903.5032}.

\bibitem[{\citenamefont{Li}(2010)}]{Li:2009rk}
\bibinfo{author}{\bibfnamefont{B.~A.} \bibnamefont{Li}},
  \bibinfo{journal}{Phys.Rev.} \textbf{\bibinfo{volume}{D81}},
  \bibinfo{pages}{114002} (\bibinfo{year}{2010}), \eprint{0912.2323}.

\bibitem[{\citenamefont{Akeroyd et~al.}(2007)\citenamefont{Akeroyd, Chen, and
  Geng}}]{Akeroyd:2007fy}
\bibinfo{author}{\bibfnamefont{A.~G.} \bibnamefont{Akeroyd}},
  \bibinfo{author}{\bibfnamefont{C.-H.} \bibnamefont{Chen}}, \bibnamefont{and}
  \bibinfo{author}{\bibfnamefont{C.-Q.} \bibnamefont{Geng}},
  \bibinfo{journal}{Phys. Rev.} \textbf{\bibinfo{volume}{D75}},
  \bibinfo{pages}{054003} (\bibinfo{year}{2007}), \eprint{hep-ph/0701012}.

\bibitem[{\citenamefont{Okubo}(1963)}]{OZI1}
\bibinfo{author}{\bibfnamefont{S.}~\bibnamefont{Okubo}},
  \bibinfo{journal}{Phys. Lett.} \textbf{\bibinfo{volume}{5}},
  \bibinfo{pages}{165} (\bibinfo{year}{1963}).

\bibitem[{\citenamefont{Zweig}(1964)}]{OZI2}
\bibinfo{author}{\bibfnamefont{G.}~\bibnamefont{Zweig}}, \bibinfo{journal}{CERN
  Report} \textbf{\bibinfo{volume}{TH-401}} (\bibinfo{year}{1964}).

\bibitem[{\citenamefont{Iizuka}(1966)}]{OZI3}
\bibinfo{author}{\bibfnamefont{I.}~\bibnamefont{Iizuka}},
  \bibinfo{journal}{Prog. Theor. Phys. Suppl.} \textbf{\bibinfo{volume}{37}},
  \bibinfo{pages}{21} (\bibinfo{year}{1966}).

\bibitem[{\citenamefont{Li and Zhao}(2008)}]{Li:2007xr}
\bibinfo{author}{\bibfnamefont{G.}~\bibnamefont{Li}} \bibnamefont{and}
  \bibinfo{author}{\bibfnamefont{Q.}~\bibnamefont{Zhao}},
  \bibinfo{journal}{Phys.Lett.} \textbf{\bibinfo{volume}{B670}},
  \bibinfo{pages}{55} (\bibinfo{year}{2008}), \eprint{0709.4639}.

\bibitem[{\citenamefont{Li and Zhao}(2011)}]{Li:2011ss}
\bibinfo{author}{\bibfnamefont{G.}~\bibnamefont{Li}} \bibnamefont{and}
  \bibinfo{author}{\bibfnamefont{Q.}~\bibnamefont{Zhao}},
  \bibinfo{journal}{Phys.Rev.} \textbf{\bibinfo{volume}{D84}},
  \bibinfo{pages}{074005} (\bibinfo{year}{2011}), \eprint{1107.2037}.

\bibitem[{\citenamefont{Xiao et~al.}(2008)\citenamefont{Xiao, Zhang, Liu, and
  Guo}}]{Xiao:2008sw}
\bibinfo{author}{\bibfnamefont{Z.-J.} \bibnamefont{Xiao}},
  \bibinfo{author}{\bibfnamefont{Z.-Q.} \bibnamefont{Zhang}},
  \bibinfo{author}{\bibfnamefont{X.}~\bibnamefont{Liu}}, \bibnamefont{and}
  \bibinfo{author}{\bibfnamefont{L.-B.} \bibnamefont{Guo}},
  \bibinfo{journal}{Phys. Rev.} \textbf{\bibinfo{volume}{D78}},
  \bibinfo{pages}{114001} (\bibinfo{year}{2008}), \eprint{0807.4265}.

\bibitem[{\citenamefont{Liu et~al.}(2009)\citenamefont{Liu, Zhang, and
  Xiao}}]{Liu:2009kz}
\bibinfo{author}{\bibfnamefont{X.}~\bibnamefont{Liu}},
  \bibinfo{author}{\bibfnamefont{Z.-Q.} \bibnamefont{Zhang}}, \bibnamefont{and}
  \bibinfo{author}{\bibfnamefont{Z.-J.} \bibnamefont{Xiao}}
  (\bibinfo{year}{2009}), \eprint{0901.0165}.

\bibitem[{\citenamefont{Schechter et~al.}(1993)\citenamefont{Schechter,
  Subbaraman, and Weigel}}]{Schechter:1992iz}
\bibinfo{author}{\bibfnamefont{J.}~\bibnamefont{Schechter}},
  \bibinfo{author}{\bibfnamefont{A.}~\bibnamefont{Subbaraman}},
  \bibnamefont{and} \bibinfo{author}{\bibfnamefont{H.}~\bibnamefont{Weigel}},
  \bibinfo{journal}{Phys.Rev.} \textbf{\bibinfo{volume}{D48}},
  \bibinfo{pages}{339} (\bibinfo{year}{1993}), \eprint{hep-ph/9211239}.

\bibitem[{\citenamefont{Kou}(2001)}]{Kou:1999tt}
\bibinfo{author}{\bibfnamefont{E.}~\bibnamefont{Kou}}, \bibinfo{journal}{Phys.
  Rev.} \textbf{\bibinfo{volume}{D63}}, \bibinfo{pages}{054027}
  (\bibinfo{year}{2001}), \eprint{hep-ph/9908214}.

\bibitem[{\citenamefont{Amsler et~al.}(2008)}]{Amsler:2008zb}
\bibinfo{author}{\bibfnamefont{C.}~\bibnamefont{Amsler}} \bibnamefont{et~al.}
  (\bibinfo{collaboration}{Particle Data Group}), \bibinfo{journal}{Phys.Lett.}
  \textbf{\bibinfo{volume}{B667}}, \bibinfo{pages}{1} (\bibinfo{year}{2008}).

\bibitem[{\citenamefont{Nakamura et~al.}(2010)}]{Nakamura:2010zzi}
\bibinfo{author}{\bibfnamefont{K.}~\bibnamefont{Nakamura}} \bibnamefont{et~al.}
  (\bibinfo{collaboration}{Particle Data Group}), \bibinfo{journal}{J.Phys.G}
  \textbf{\bibinfo{volume}{G37}}, \bibinfo{pages}{075021}
  (\bibinfo{year}{2010}).

\bibitem[{\citenamefont{Ablikim et~al.}(2011)}]{Collaboration:2011ii}
\bibinfo{author}{\bibfnamefont{M.}~\bibnamefont{Ablikim}} \bibnamefont{et~al.}
  (\bibinfo{year}{2011}), \eprint{1111.0398}.

\bibitem[{\citenamefont{Kwong et~al.}(1988)\citenamefont{Kwong, Mackenzie,
  Rosenfeld, and Rosner}}]{Kwong:1987ak}
\bibinfo{author}{\bibfnamefont{W.}~\bibnamefont{Kwong}},
  \bibinfo{author}{\bibfnamefont{P.~B.} \bibnamefont{Mackenzie}},
  \bibinfo{author}{\bibfnamefont{R.}~\bibnamefont{Rosenfeld}},
  \bibnamefont{and} \bibinfo{author}{\bibfnamefont{J.~L.}
  \bibnamefont{Rosner}}, \bibinfo{journal}{Phys.Rev.}
  \textbf{\bibinfo{volume}{D37}}, \bibinfo{pages}{3210} (\bibinfo{year}{1988}).

\bibitem[{\citenamefont{Close and Zhao}(2005)}]{Close:2005vf}
\bibinfo{author}{\bibfnamefont{F.~E.} \bibnamefont{Close}} \bibnamefont{and}
  \bibinfo{author}{\bibfnamefont{Q.}~\bibnamefont{Zhao}},
  \bibinfo{journal}{Phys.Rev.} \textbf{\bibinfo{volume}{D71}},
  \bibinfo{pages}{094022} (\bibinfo{year}{2005}), \eprint{hep-ph/0504043}.

\bibitem[{\citenamefont{Guo et~al.}(2011)\citenamefont{Guo, Ma, and
  Chao}}]{Guo:2011tz}
\bibinfo{author}{\bibfnamefont{H.-K.} \bibnamefont{Guo}},
  \bibinfo{author}{\bibfnamefont{Y.-Q.} \bibnamefont{Ma}}, \bibnamefont{and}
  \bibinfo{author}{\bibfnamefont{K.-T.} \bibnamefont{Chao}},
  \bibinfo{journal}{Phys. Rev.} \textbf{\bibinfo{volume}{D83}},
  \bibinfo{pages}{114038} (\bibinfo{year}{2011}), \eprint{1104.3138}.

\bibitem[{\citenamefont{Charng et~al.}(2006)\citenamefont{Charng, Kurimoto, and
  Li}}]{Charng:2006zj}
\bibinfo{author}{\bibfnamefont{Y.-Y.} \bibnamefont{Charng}},
  \bibinfo{author}{\bibfnamefont{T.}~\bibnamefont{Kurimoto}}, \bibnamefont{and}
  \bibinfo{author}{\bibfnamefont{H.-n.} \bibnamefont{Li}},
  \bibinfo{journal}{Phys.Rev.} \textbf{\bibinfo{volume}{D74}},
  \bibinfo{pages}{074024} (\bibinfo{year}{2006}), \eprint{hep-ph/0609165}.

\bibitem[{\citenamefont{Yuan and Chao}(1997)}]{Yuan:1997ts}
\bibinfo{author}{\bibfnamefont{F.}~\bibnamefont{Yuan}} \bibnamefont{and}
  \bibinfo{author}{\bibfnamefont{K.-T.} \bibnamefont{Chao}},
  \bibinfo{journal}{Phys. Rev.} \textbf{\bibinfo{volume}{D56}},
  \bibinfo{pages}{R2495} (\bibinfo{year}{1997}), \eprint{hep-ph/9706294}.

\bibitem[{\citenamefont{Wu and Huang}(2011)}]{Wu:2011gf}
\bibinfo{author}{\bibfnamefont{X.-G.} \bibnamefont{Wu}} \bibnamefont{and}
  \bibinfo{author}{\bibfnamefont{T.}~\bibnamefont{Huang}},
  \bibinfo{journal}{Phys.Rev.} \textbf{\bibinfo{volume}{D84}},
  \bibinfo{pages}{074011} (\bibinfo{year}{2011}), \bibinfo{note}{18 pages, 8
  figures and 1 table. Reference updated. To be published in Phys.Rev.D},
  \eprint{1106.4365}.

\bibitem[{\citenamefont{Franz et~al.}(2000)\citenamefont{Franz, Polyakov, and
  Goeke}}]{Franz:2000ee}
\bibinfo{author}{\bibfnamefont{M.}~\bibnamefont{Franz}},
  \bibinfo{author}{\bibfnamefont{M.~V.} \bibnamefont{Polyakov}},
  \bibnamefont{and} \bibinfo{author}{\bibfnamefont{K.}~\bibnamefont{Goeke}},
  \bibinfo{journal}{Phys.Rev.} \textbf{\bibinfo{volume}{D62}},
  \bibinfo{pages}{074024} (\bibinfo{year}{2000}), \eprint{hep-ph/0002240}.

\bibitem[{\citenamefont{Ali and Greub}(1998)}]{Ali:1997nh}
\bibinfo{author}{\bibfnamefont{A.}~\bibnamefont{Ali}} \bibnamefont{and}
  \bibinfo{author}{\bibfnamefont{C.}~\bibnamefont{Greub}},
  \bibinfo{journal}{Phys.Rev.} \textbf{\bibinfo{volume}{D57}},
  \bibinfo{pages}{2996} (\bibinfo{year}{1998}), \eprint{hep-ph/9707251}.

\bibitem[{\citenamefont{Feldmann and Kroll}(2000)}]{Feldmann:2000hs}
\bibinfo{author}{\bibfnamefont{T.}~\bibnamefont{Feldmann}} \bibnamefont{and}
  \bibinfo{author}{\bibfnamefont{P.}~\bibnamefont{Kroll}},
  \bibinfo{journal}{Phys.Rev.} \textbf{\bibinfo{volume}{D62}},
  \bibinfo{pages}{074006} (\bibinfo{year}{2000}), \eprint{hep-ph/0003096}.

\bibitem[{\citenamefont{Gerard and Kou}(2006)}]{Gerard:2006ch}
\bibinfo{author}{\bibfnamefont{J.-M.} \bibnamefont{Gerard}} \bibnamefont{and}
  \bibinfo{author}{\bibfnamefont{E.}~\bibnamefont{Kou}},
  \bibinfo{journal}{Phys.Rev.Lett.} \textbf{\bibinfo{volume}{97}},
  \bibinfo{pages}{261804} (\bibinfo{year}{2006}), \eprint{hep-ph/0609300}.

\bibitem[{\citenamefont{Mathieu and Vento}(2010)}]{Mathieu:2009sg}
\bibinfo{author}{\bibfnamefont{V.}~\bibnamefont{Mathieu}} \bibnamefont{and}
  \bibinfo{author}{\bibfnamefont{V.}~\bibnamefont{Vento}},
  \bibinfo{journal}{Phys.Rev.} \textbf{\bibinfo{volume}{D81}},
  \bibinfo{pages}{034004} (\bibinfo{year}{2010}), \eprint{0910.0212}.

\bibitem[{HFA()}]{HFAG}
\bibinfo{note}{Heavy Flavor Averaging Group,
  http://www.slac.stanford.edu/xorg/hfag}.

\bibitem[{\citenamefont{Godfrey and Isgur}(1985)}]{Godfrey:1985xj}
\bibinfo{author}{\bibfnamefont{S.}~\bibnamefont{Godfrey}} \bibnamefont{and}
  \bibinfo{author}{\bibfnamefont{N.}~\bibnamefont{Isgur}},
  \bibinfo{journal}{Phys.Rev.} \textbf{\bibinfo{volume}{D32}},
  \bibinfo{pages}{189} (\bibinfo{year}{1985}).

\bibitem[{\citenamefont{Barnes et~al.}(2005)\citenamefont{Barnes, Godfrey, and
  Swanson}}]{Barnes:2005pb}
\bibinfo{author}{\bibfnamefont{T.}~\bibnamefont{Barnes}},
  \bibinfo{author}{\bibfnamefont{S.}~\bibnamefont{Godfrey}}, \bibnamefont{and}
  \bibinfo{author}{\bibfnamefont{E.~S.} \bibnamefont{Swanson}},
  \bibinfo{journal}{Phys.Rev.} \textbf{\bibinfo{volume}{D72}},
  \bibinfo{pages}{054026} (\bibinfo{year}{2005}), \eprint{hep-ph/0505002}.

\bibitem[{\citenamefont{Mitchell et~al.}(2009)}]{:2008fb}
\bibinfo{author}{\bibfnamefont{R.}~\bibnamefont{Mitchell}} \bibnamefont{et~al.}
  (\bibinfo{collaboration}{CLEO Collaboration}),
  \bibinfo{journal}{Phys.Rev.Lett.} \textbf{\bibinfo{volume}{102}},
  \bibinfo{pages}{011801} (\bibinfo{year}{2009}), \eprint{0805.0252}.

\bibitem[{\citenamefont{Eichten et~al.}(1975)\citenamefont{Eichten, Gottfried,
  Kinoshita, Kogut, Lane et~al.}}]{Eichten:1974af}
\bibinfo{author}{\bibfnamefont{E.}~\bibnamefont{Eichten}},
  \bibinfo{author}{\bibfnamefont{K.}~\bibnamefont{Gottfried}},
  \bibinfo{author}{\bibfnamefont{T.}~\bibnamefont{Kinoshita}},
  \bibinfo{author}{\bibfnamefont{J.~B.} \bibnamefont{Kogut}},
  \bibinfo{author}{\bibfnamefont{K.~D.} \bibnamefont{Lane}},
  \bibnamefont{et~al.}, \bibinfo{journal}{Phys.Rev.Lett.}
  \textbf{\bibinfo{volume}{34}}, \bibinfo{pages}{369} (\bibinfo{year}{1975}).

\bibitem[{\citenamefont{Eichten et~al.}(1976)\citenamefont{Eichten, Gottfried,
  Kinoshita, Lane, and Yan}}]{Eichten:1975ag}
\bibinfo{author}{\bibfnamefont{E.}~\bibnamefont{Eichten}},
  \bibinfo{author}{\bibfnamefont{K.}~\bibnamefont{Gottfried}},
  \bibinfo{author}{\bibfnamefont{T.}~\bibnamefont{Kinoshita}},
  \bibinfo{author}{\bibfnamefont{K.~D.} \bibnamefont{Lane}}, \bibnamefont{and}
  \bibinfo{author}{\bibfnamefont{T.-M.} \bibnamefont{Yan}},
  \bibinfo{journal}{Phys.Rev.Lett.} \textbf{\bibinfo{volume}{36}},
  \bibinfo{pages}{500} (\bibinfo{year}{1976}).

\bibitem[{\citenamefont{Eichten et~al.}(1978)\citenamefont{Eichten, Gottfried,
  Kinoshita, Lane, and Yan}}]{Eichten:1978tg}
\bibinfo{author}{\bibfnamefont{E.}~\bibnamefont{Eichten}},
  \bibinfo{author}{\bibfnamefont{K.}~\bibnamefont{Gottfried}},
  \bibinfo{author}{\bibfnamefont{T.}~\bibnamefont{Kinoshita}},
  \bibinfo{author}{\bibfnamefont{K.~D.} \bibnamefont{Lane}}, \bibnamefont{and}
  \bibinfo{author}{\bibfnamefont{T.-M.} \bibnamefont{Yan}},
  \bibinfo{journal}{Phys.Rev.} \textbf{\bibinfo{volume}{D17}},
  \bibinfo{pages}{3090} (\bibinfo{year}{1978}).

\bibitem[{\citenamefont{Eichten et~al.}(1980)\citenamefont{Eichten, Gottfried,
  Kinoshita, Lane, and Yan}}]{Eichten:1979ms}
\bibinfo{author}{\bibfnamefont{E.}~\bibnamefont{Eichten}},
  \bibinfo{author}{\bibfnamefont{K.}~\bibnamefont{Gottfried}},
  \bibinfo{author}{\bibfnamefont{T.}~\bibnamefont{Kinoshita}},
  \bibinfo{author}{\bibfnamefont{K.~D.} \bibnamefont{Lane}}, \bibnamefont{and}
  \bibinfo{author}{\bibfnamefont{T.-M.} \bibnamefont{Yan}},
  \bibinfo{journal}{Phys.Rev.} \textbf{\bibinfo{volume}{D21}},
  \bibinfo{pages}{203} (\bibinfo{year}{1980}).

\bibitem[{\citenamefont{Ebert et~al.}(2003)\citenamefont{Ebert, Faustov, and
  Galkin}}]{Ebert:2002pp}
\bibinfo{author}{\bibfnamefont{D.}~\bibnamefont{Ebert}},
  \bibinfo{author}{\bibfnamefont{R.~N.} \bibnamefont{Faustov}},
  \bibnamefont{and} \bibinfo{author}{\bibfnamefont{V.~O.}
  \bibnamefont{Galkin}}, \bibinfo{journal}{Phys.Rev.}
  \textbf{\bibinfo{volume}{D67}}, \bibinfo{pages}{014027}
  (\bibinfo{year}{2003}), \eprint{hep-ph/0210381}.

\bibitem[{\citenamefont{Eichten et~al.}(2008)\citenamefont{Eichten, Godfrey,
  Mahlke, and Rosner}}]{Eichten:2007qx}
\bibinfo{author}{\bibfnamefont{E.}~\bibnamefont{Eichten}},
  \bibinfo{author}{\bibfnamefont{S.}~\bibnamefont{Godfrey}},
  \bibinfo{author}{\bibfnamefont{H.}~\bibnamefont{Mahlke}}, \bibnamefont{and}
  \bibinfo{author}{\bibfnamefont{J.~L.} \bibnamefont{Rosner}},
  \bibinfo{journal}{Rev.Mod.Phys.} \textbf{\bibinfo{volume}{80}},
  \bibinfo{pages}{1161} (\bibinfo{year}{2008}), \eprint{hep-ph/0701208}.

\bibitem[{\citenamefont{Brambilla et~al.}(2006)\citenamefont{Brambilla, Jia,
  and Vairo}}]{Brambilla:2005zw}
\bibinfo{author}{\bibfnamefont{N.}~\bibnamefont{Brambilla}},
  \bibinfo{author}{\bibfnamefont{Y.}~\bibnamefont{Jia}}, \bibnamefont{and}
  \bibinfo{author}{\bibfnamefont{A.}~\bibnamefont{Vairo}},
  \bibinfo{journal}{Phys.Rev.} \textbf{\bibinfo{volume}{D73}},
  \bibinfo{pages}{054005} (\bibinfo{year}{2006}), \eprint{hep-ph/0512369}.

\bibitem[{\citenamefont{Brambilla et~al.}(2004)}]{Brambilla:2004wf}
\bibinfo{author}{\bibfnamefont{N.}~\bibnamefont{Brambilla}}
  \bibnamefont{et~al.} (\bibinfo{collaboration}{Quarkonium Working Group})
  (\bibinfo{year}{2004}), \bibinfo{note}{published as CERN Yellow Report,
  CERN-2005-005, Geneva: CERN, 2005. -487 p.}, \eprint{hep-ph/0412158}.

\bibitem[{\citenamefont{Dudek et~al.}(2006)\citenamefont{Dudek, Edwards, and
  Richards}}]{Dudek:2006ej}
\bibinfo{author}{\bibfnamefont{J.~J.} \bibnamefont{Dudek}},
  \bibinfo{author}{\bibfnamefont{R.~G.} \bibnamefont{Edwards}},
  \bibnamefont{and} \bibinfo{author}{\bibfnamefont{D.~G.}
  \bibnamefont{Richards}}, \bibinfo{journal}{Phys.Rev.}
  \textbf{\bibinfo{volume}{D73}}, \bibinfo{pages}{074507}
  (\bibinfo{year}{2006}), \eprint{hep-ph/0601137}.

\bibitem[{\citenamefont{Dudek et~al.}(2009)\citenamefont{Dudek, Edwards, and
  Thomas}}]{Dudek:2009kk}
\bibinfo{author}{\bibfnamefont{J.~J.} \bibnamefont{Dudek}},
  \bibinfo{author}{\bibfnamefont{R.~G.} \bibnamefont{Edwards}},
  \bibnamefont{and} \bibinfo{author}{\bibfnamefont{C.~E.}
  \bibnamefont{Thomas}}, \bibinfo{journal}{Phys.Rev.}
  \textbf{\bibinfo{volume}{D79}}, \bibinfo{pages}{094504}
  (\bibinfo{year}{2009}), \eprint{0902.2241}.

\bibitem[{\citenamefont{Gerard and Trine}(2004)}]{Gerard:2004je}
\bibinfo{author}{\bibfnamefont{J.~M.} \bibnamefont{Gerard}} \bibnamefont{and}
  \bibinfo{author}{\bibfnamefont{S.}~\bibnamefont{Trine}},
  \bibinfo{journal}{Phys. Rev.} \textbf{\bibinfo{volume}{D69}},
  \bibinfo{pages}{113005} (\bibinfo{year}{2004}), \eprint{hep-ph/0402158}.

\bibitem[{\citenamefont{Wu et~al.}(2011)\citenamefont{Wu, Liu, Zhao, and
  Zou}}]{Wu:2011yx}
\bibinfo{author}{\bibfnamefont{J.-J.} \bibnamefont{Wu}},
  \bibinfo{author}{\bibfnamefont{X.-H.} \bibnamefont{Liu}},
  \bibinfo{author}{\bibfnamefont{Q.}~\bibnamefont{Zhao}}, \bibnamefont{and}
  \bibinfo{author}{\bibfnamefont{B.-S.} \bibnamefont{Zou}}
  (\bibinfo{year}{2011}), \eprint{1108.3772}.

\end{thebibliography}
\end{document}